\documentclass[pdflatex,sn-mathphys-num]{sn-jnl}

\usepackage{graphicx}
\usepackage{multirow}
\usepackage{amsmath,amssymb,amsfonts}
\usepackage{amsthm}
\usepackage{mathrsfs}
\usepackage[title]{appendix}
\usepackage{xcolor}
\usepackage{textcomp}
\usepackage{manyfoot}
\usepackage{booktabs}
\usepackage{algorithm}
\usepackage{algorithmicx}
\usepackage{algpseudocode}
\usepackage{listings}
\DeclareGraphicsExtensions{.pdf,.png}
\usepackage{caption}
\usepackage{subcaption}
\usepackage{soul}
\usepackage{adjustbox}
\usepackage{rotating}
\usepackage{glossaries}

\usepackage{xurl}
\usepackage{tabularx}
\usepackage[group-separator={,}]{siunitx}
\usepackage{orcidlink}

\theoremstyle{thmstyleone}

\theoremstyle{thmstyletwo}

\theoremstyle{thmstylethree}

\begin{document}

\title[Article Title]{Workplace dependence in urban economies}

\author*[1,2]{\fnm{Zs\'{o}fia} \sur{Z\'{a}dor}}

\author[3,4,5]{\fnm{Bal\'{a}zs} \sur{Lengyel}}

\author[1,6]{\fnm{Riccardo} \sur{Di Clemente}}

\small\affil*[1]{\orgdiv{Complex Connections Lab, Network Science Institute}, \orgname{Northeastern University London}, \orgaddress{\street{St Katharine's Way}, \city{London}, \postcode{E1W 1LP}, \country{UK}}}

\small\affil[2]{\orgdiv{Faculty of Business}, \orgname{Northeastern University London}, \orgaddress{\street{St Katharine's Way}, \city{London}, \postcode{E1W 1LP}, \country{UK}}}

\small\affil[3]{\orgdiv{MTA-ELTE Agglomeration, Networks, and Innovation Momentum Research Group}, \orgname{ELTE Centre for Economic and Regional Studies}, \orgaddress{\street{Tóth Kálman Street 4}, \city{Budapest}, \postcode{1097}, \country{Hungary}}}

\small\affil[4]{\orgdiv{ANETI Lab, Corvinus Institute for Advanced Studies}, \orgname{Corvinus University of Budapest}, \orgaddress{\street{Közraktár street 4-6}, \city{Budapest}, \postcode{1093}, \country{Hungary}}}

\small\affil[5]{\orgdiv{Institute for Data Analytics and Information Systems}, \orgname{Corvinus University of Budapest}, \orgaddress{\street{Közraktár street 4-6}, \city{Budapest}, \postcode{1093}, \country{Hungary}}}

\small\affil[6]{\orgdiv{ISI Foundation}, \orgaddress{\street{Chisola n. 5}, \city{Torino}, \postcode{10126}, \country{Italy}}}

\abstract{Remote work has fundamentally reshaped urban economic life, 
and the spatial organisation of activity across cities. However, access to flexible work is distributed unevenly across industries, income groups, and genders, creating disparities in health risks, social mixing, and economic opportunity.
Understanding where workplace dependence (WPD) is concentrated is therefore important, yet its distribution across urban areas remains poorly understood.
Here we pair fine-grained hourly population data with detailed company records in a large European city to examine how location, industry composition, and socio-economic characteristics shape physical workplace attendance. 
By comparing workplace activity during periods of low versus high COVID-19 restrictions, we identify the determinants of WPD.
We find that while industry and firm productivity are key drivers, the relationship between WPD, income, and gender is highly contingent on distance from the city center. Near the centre, female-majority and income-diverse locations show the highest WPD, consistent with a residual, place-bound service workforce.
These findings reveal a spatially contingent 'service trap' at the urban core, extending remote-work inequalities beyond individuals to the urban ecosystem as a whole.}

\keywords{remote work, spatial inequality, urban structure, industry differences, socio-economic differences, mobile phone data, administrative firm data}

\maketitle

\section{Introduction}\label{sec1}
The COVID-19 pandemic fundamentally altered urban economic activity by forcing a rapid, unplanned experiment in remote work \cite{barrero_evolution_2023, bick_work_2023, barrero_why_2021}, triggering persistent shifts in the spatio-temporal dimensions of human mobility \cite{santana2023covid}. This transition was unequal. High-skilled, high-income workers in male-dominated industries were far more likely to benefit \cite{dingel_how_2020, koren2020business, adams2022work, minkus2022significance, huang_staying_2022}. Those whose jobs required physical presence not only lost flexibility but also bore greater health risks during the pandemic~\cite{lunde2022relationship, gozzi_estimating_2021}. Access to remote work varies by income and occupation, and preferences for the remote option differ by gender and caring responsibilities~\cite{aksoy_working_2022}. The distributional consequences extend beyond those who gain the option: when high-skilled, high-income workers leave city centres, the low-skilled consumer service workers who depended on their proximity suffer the consequences \cite{althoff_geography_2022}.

This reallocation of activity was geographically concentrated: while peripheral residential areas benefited from increased local spending~\cite{ramani_how_2024}, the absence of workers has led to dramatic drop of lease prices in the office real estate market \cite{gupta2026work}, shocked the services around workplaces \cite{yabe_behaviour-based_2024}, and have decreased the potential of social mixing in the city \cite{yabe_behavioral_2023}. This spatial redistribution of activity is described as the ``Donut Effect''~\cite{ramani_how_2024, yabe_behaviour-based_2024}. Our analysis suggests that this emptying did not happen in a uniform manner. It followed socio-economic filtering. The spatial logic of the city shifts towards essential service maintenance, which affects not only who remains at workplaces, but also the redistribution of social encounters.

These disparities are further sharpened by the fact that workplace location is a key driver of economic and social opportunity. Urban centres concentrate industries that have high productivity-land ratio \cite{alonso1964location} and high-skilled workers \cite{moretti2012new, glaeser2001cities, althoff_geography_2022}, but at the same time, are also arenas of social mixing \cite{zhong2017revealing,moro_mobility_2021, juhasz2023amenity} establishing diverse environments that makes cities thrive \cite{jacobs_death_1972}. Inequalities in such mixing patterns have been widely demonstrated. For example, spatial sorting limits the workplace opportunities of low-status groups living in the urban periphery \cite{brueckner1999central, zhang2020dynamics}, while women tend to commute shorter distances \cite{white1986sex}, contributing to spatially segmented labour markets \cite{hanson1988spatial}. These segmented markets are reinforced by gendered mobility constraints; women’s activity patterns are often more spatially restricted and temporally distinct from men’s \cite{macedo_differences_2022, collins2024spatiotemporal}. Such constraints mean that the spatial reorganisation triggered by remote work may reinforce existing inequalities.

Although the potential for remote working is well documented, we lack an understanding of \textit{workplace dependence} (WPD), which captures the residual physical presence necessary for essential urban functions -- the complement of remote-work potential. This pattern is concentrated in essential services and is distributed unequally across socio-economic groups. However, the methods developed to study remote work are not designed to observe WPD directly. Existing research has either inferred remote work potential from occupational classifications~\cite{dingel_how_2020, yasenov_who_2020}, relied on surveys that miss spatial and firm-level dynamics~\cite{aksoy_working_2022, caros_need_2024}, or modelled hypothetical urban equilibria rather than observing realised attendance behaviour~\cite{delventhal_jue_2022, monte_remote_2023}. Where mobility data have been used to measure attendance directly~\cite{gibbs_harnessing_2023}, studies have operated at aggregate administrative geographies and without linkage to firm-level records. As a result, we do not yet know how socio-economic and spatial characteristics jointly determine which workplaces---and which workers---maintained physical presence when remote work became broadly available.

This study investigates how spatial location, industry composition, and socio-economic characteristics shape WPD during a period when remote work became widely feasible. We focus on the spatial residual: the workers and places that maintained physical presence. Using longitudinal geolocated data from 1.4 million mobile devices in Budapest, Hungary, we measure WPD as the degree to which workers continued to attend their workplaces between an unrestricted ``opening period'' (September--October 2020) and a strict ``curfew period'' (November--December 2020). Budapest provides a particularly informative empirical setting: its dense, monocentric structure and the abruptness of the November 2020 curfew create an intervention that isolates structural workplace dependence from elective commuting, allowing us to observe revealed remote-work capacity at sub-kilometre resolution. By linking these mobility patterns with administrative company records from OPTEN, we can identify which places and sectors continued to rely on a physical presence, and which shifted towards remote working arrangements. This combination of factors has not previously been applied to answer this question.

We find that industry and company composition are important determinants of workplace persistence: locations with a high diversity of firms and a strong commerce presence exhibit greater continuity of on-site attendance than knowledge-intensive service sectors.
Beyond these structural factors, three characteristics further differentiate patterns of WPD. Locations farther from the city centre, and those with female-majority daytime populations, show the strongest persistence across industries. This spatial and socio-economic patterning reveals that workplace dependence closely followed pre-existing lines of disadvantage, confirming that those excluded from the remote work option were often the most vulnerable groups in the urban economy. In central locations, female-majority and income-diverse workplaces show higher WPD, consistent with a concentration of face-to-face service workers who remained tied to the urban core to serve a now-remote-capable workforce. In peripheral locations, this signal weakens, and physical presence is driven more by industrial land-use patterns than by workforce demographics. Together, these findings reveal a `service trap' in the urban core: proximity to the city centre compounds workplace dependence, adding a geographic dimension to a labour market inequality that already falls most heavily on those with the least capacity to work remotely.

\section{Results}\label{sec2}

To examine how spatial location, industry composition, and socio-economic characteristics shape workplace dependence, we combine hourly mobile phone population counts with administrative company records across Budapest's H3 hexagonal grid, comparing working activity between an unrestricted opening period (September–October 2020) and a strict curfew period (November–December 2020). Although mobile phone data has been widely used to track urban mobility and socio-economic segregation, there has been little attempt to link this data with administrative company records. This allows us to characterise the industry composition and workforce demographics of locations. Temporal population density data for 1.4 million devices was provided by Hungarian Telekom, which identified work and home locations based on frequent visitation patterns during working and resting hours. The company also stratified the data based on behaviour and contract information. We aggregate company administrative records obtained from OPTEN between 2018 and 2020 at the H3 hexagon level, matching them with population density information. Using data from 2018 enables us to understand companies before the impact of the supply chain and employment shocks caused by the pandemic. Of the 178,374 registered companies in Budapest, 29,409 have more than three employees; these are matched to 8,861 of the 34,516 H3 hexagons available across the city. The matched locations contain on average three companies and approximately 89 employees per hexagon, though the distribution is strongly right-skewed, with a small number of hexagons concentrating very high levels of employment.

\subsection{Temporal Rhythms and Functional Validation}\label{subsec:rhythms}
\begin{figure}[th]
    \centering
    \begin{subfigure}[t]{\linewidth}
        \includegraphics[width=\linewidth]{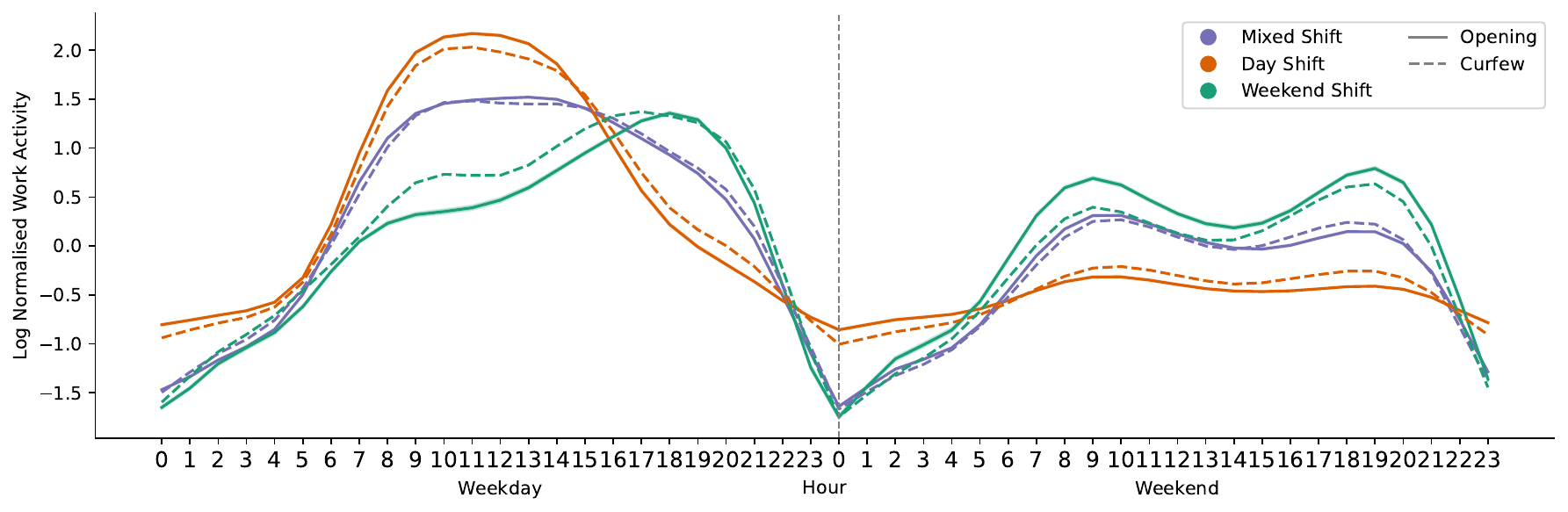}
        \caption{}
        \label{fig:rhythm_work_clust}
    \end{subfigure}

    \begin{subfigure}[t]{0.49\linewidth}
        \centering
        \includegraphics[height=5.3cm]{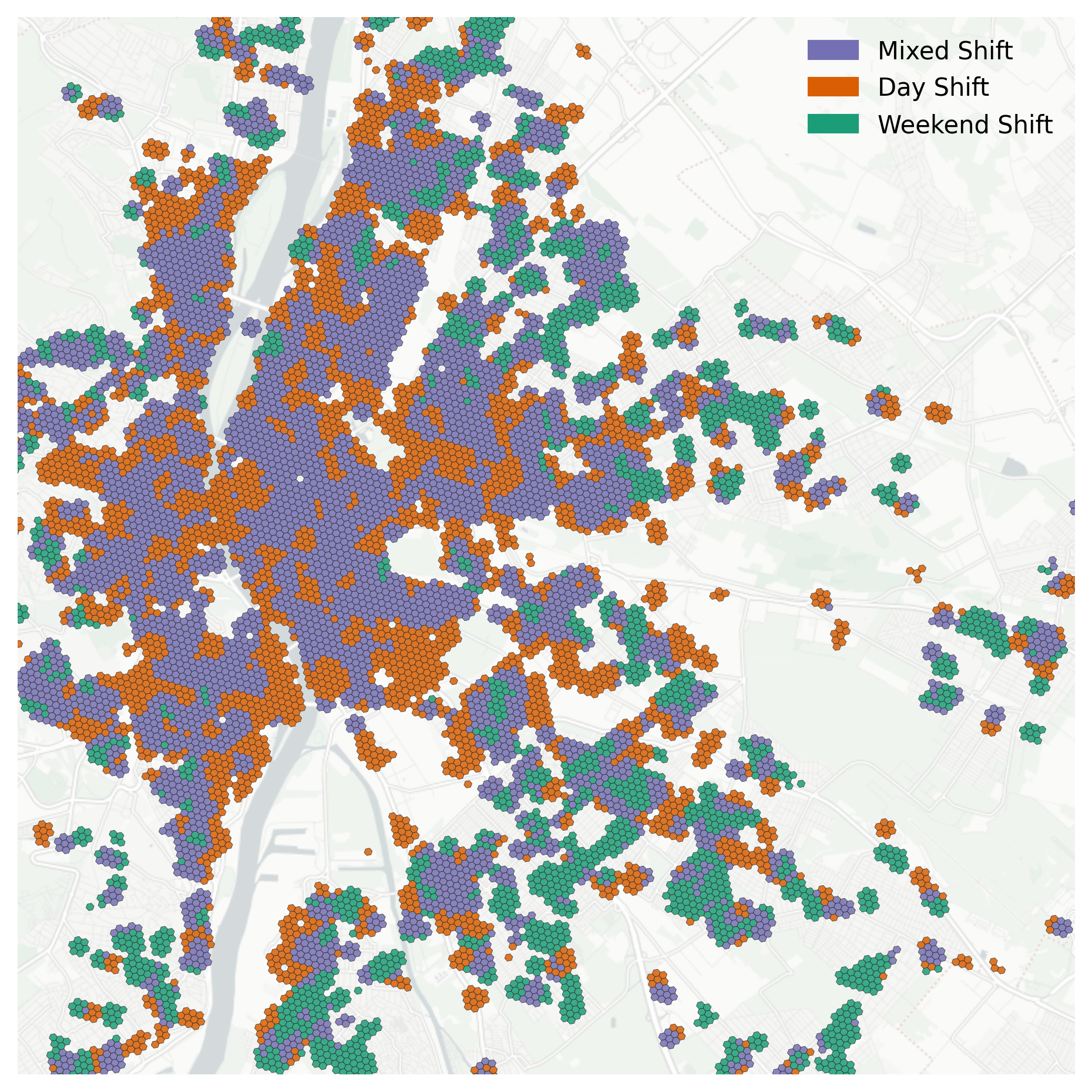}
        \captionsetup{position=bottom,justification=centering}
        \caption{}
        \label{fig:work_map}
    \end{subfigure}  
    \hfill
    \begin{subfigure}[t]{0.49\linewidth}
        \centering
        \includegraphics[height=5.3cm]{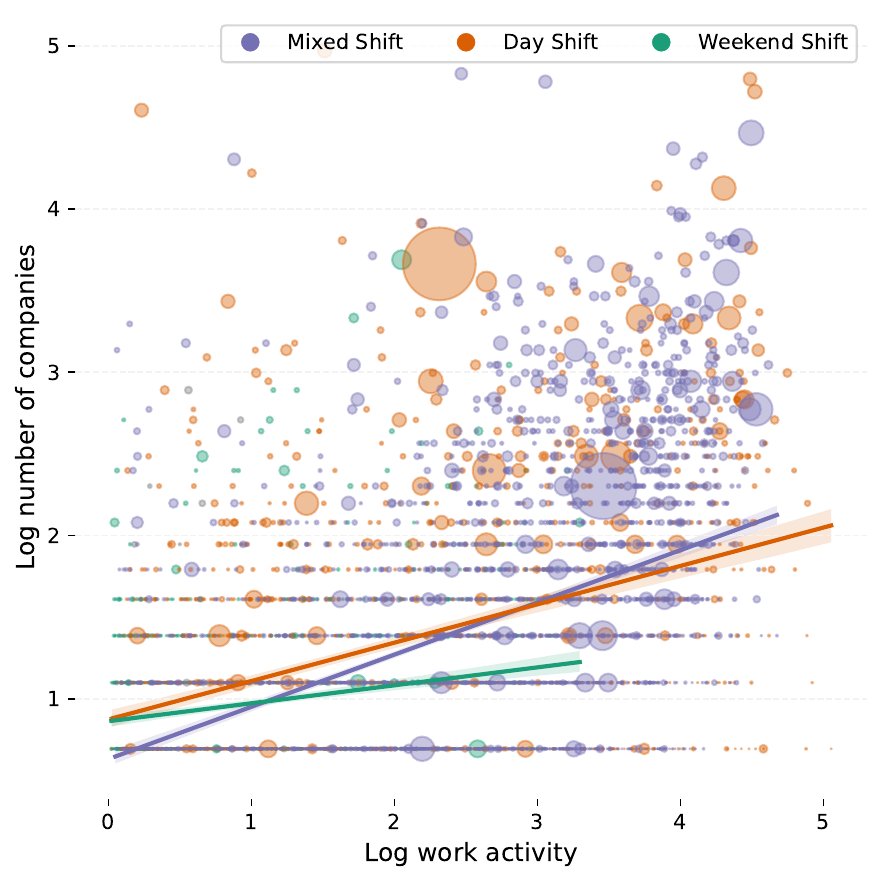}
        \captionsetup{position=bottom,justification=centering}
        \caption{}
        \label{fig:work_teaor}
    \end{subfigure}

    \caption{
        \textbf{The Interaction of working dynamics and company statistics}
        (\textbf{\subref{fig:rhythm_work_clust}}) The average work activity of the clusters during Opening and Curfew time.
        (\textbf{\subref{fig:work_map}}) Snapshot of Budapest's working activity clustered.      
        (\textbf{\subref{fig:work_teaor}}) Scatter plot of the log working activity against the log number of companies. 
        Colours represent the 3 working clusters; size = average employees per location.
    }

    \label{fig:figure}
    
\end{figure}

To characterise the spatial and temporal heterogeneity of workplace activity before modelling its determinants, we apply K-means clustering to the working activity sequences. This approach allows us to identify dominant patterns of workplace presence directly from the data, capturing distinctions in timing, intensity, and the day-of-week structure. Applying K-means clustering to the log-transformed, Gaussian-smoothed, and standardised 48-hour working activity sequence during the Opening period, we identify three clusters with distinct temporal trends (Figure~\ref{fig:rhythm_work_clust}). The \textit{"Day shift"} cluster aligns with traditional office schedules, the \textit{"Mixed shift"} cluster corresponds to multi-use spaces with extended working hours, including weekends, and the \textit{"Weekend shift"} cluster is concentrated on the urban periphery and characterised by non-standard activity patterns, such as nighttime and weekend work and increased activity during the mobility-restricted “curfew” period. 

Although industrial areas on the periphery in the \textit{Weekend shift} cluster display consistently different working patterns, there is considerable variation in activity even within central zones (Figure~\ref{fig:work_map}). Spatially, \textit{Day} and \textit{Mixed shift} clusters exhibit similar characteristics, but qualitative analysis reveals important distinctions: large commercial and institutional hubs with higher remote work adoption (\textit{Day shift}) contrast with mixed-use zones that display more diverse activity rhythms (\textit{Mixed shift}).   

To validate that the data on mobile phone usage captures genuine economic activity, we examine the log-linear relationship between working activity and the number of companies per location, as sourced from OPTEN (Figure~\ref{fig:work_teaor}). We find a strong positive association ($r = 0.475$, $R^2 = 0.226$, $p < 0.001$), confirming the concurrent validity of the mobile phone signal and the administrative company records. Cluster membership moderates this relationship: \textit{Mixed shift} locations show the steepest slope, reflecting their concentration of many firms, while \textit{Weekend shift} locations show a weaker association, consistent with their more heterogeneous economic activity (\textit{Weekend Shift}: $r = 0.235$, $R^2 = 0.055$, $p < 0.001$; \textit{Day Shift}: $r = 0.411$, $R^2 = 0.169$, $p < 0.001$; \textit{Mixed Shift}: $r = 0.493$, $R^2 = 0.243$, $p < 0.001$).

\begin{figure}[]
    \centering
    \begin{subfigure}[t]{\linewidth}
        \centering
        \includegraphics[width=0.99\linewidth]{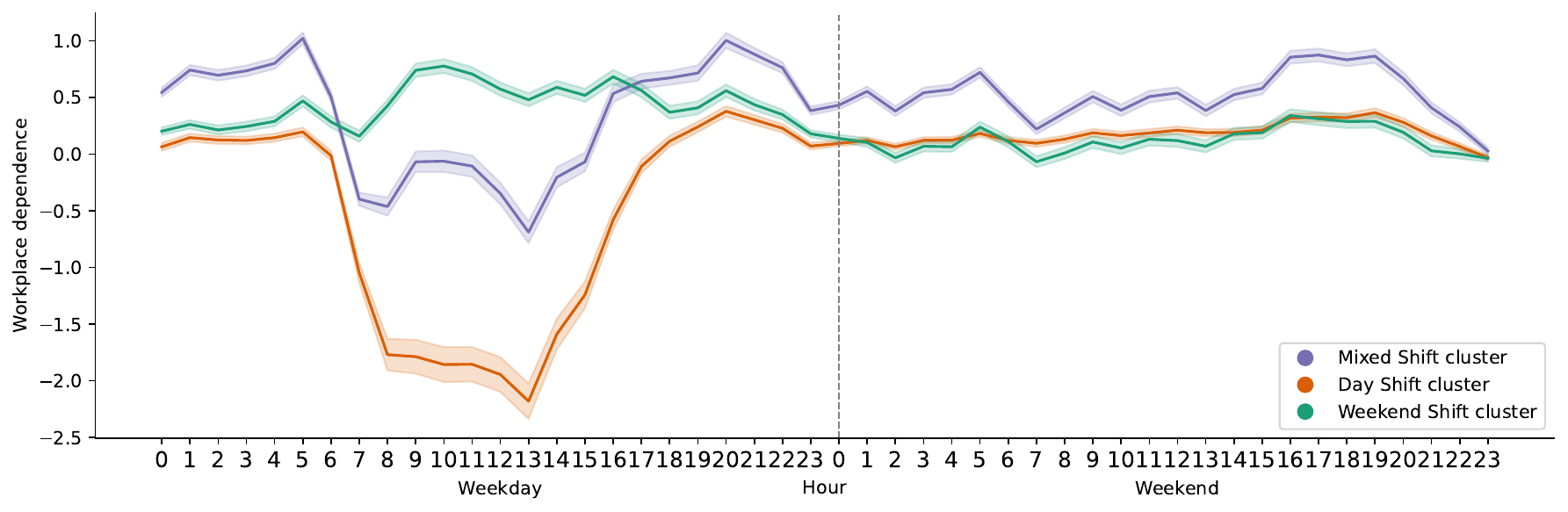}
        \caption{}
        \label{fig:change_work_activity}
    \end{subfigure}
    \vspace{0.5em}
    \begin{subfigure}[t]{\linewidth}
        \centering
        \includegraphics[width=0.6\linewidth]{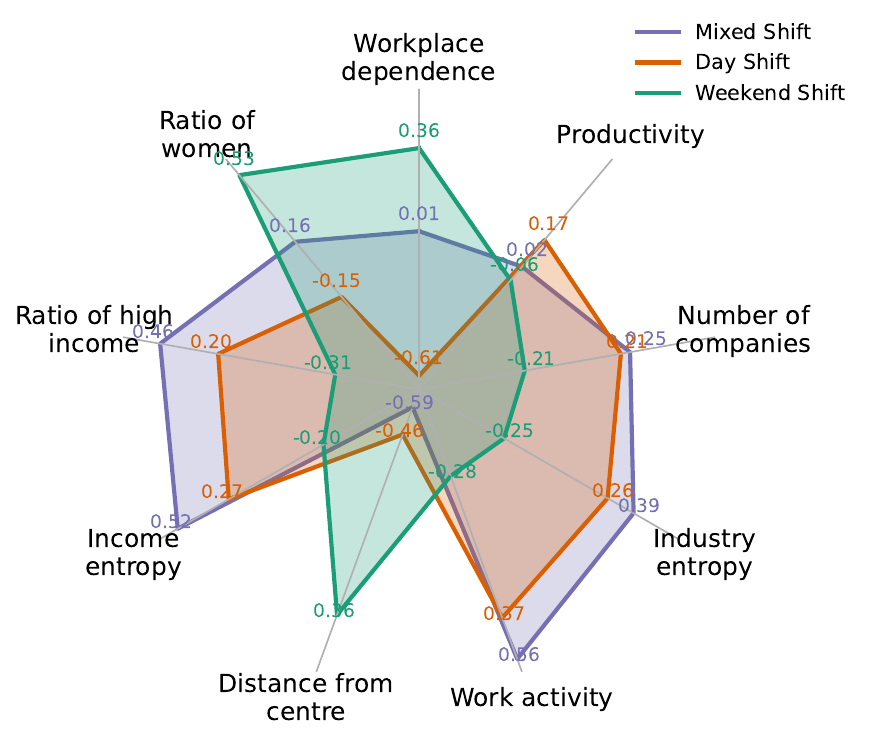}
        \caption{}
        \label{fig:radar_cluster}
    \end{subfigure}
    \caption{
        \textbf{Clustering explains variation in the variables}
        (\textbf{\subref{fig:change_work_activity}}) Change in workplace dependence by clusters.
        (\textbf{\subref{fig:radar_cluster}}) Comparison of the mean socio-economic and company variables within each working cluster. The colour of the radar plot identifies the cluster.
    }
    \label{fig:clustering_variables}
\end{figure}

\subsection{Differential Workplace Dependence}\label{subsec:persistence}
To quantify the WPD, we study the change in working activity during the COVID-19 pandemic. The introduction of the 8PM curfew on 11 November 2020 led to a decline in on-site attendance across Budapest, though this change was not uniform throughout the city. We estimate WPD for each location $i$ as $\text{workplace dependence}_i = N{\text{curfew}_i} - N_{\text{open}_i},$
where $N_{\text{curfew}_i}$ and $N_{\text{open}_i}$ denote the number of workers present during working hours at location $i$ during the Curfew and Opening periods respectively. Negative values indicate a decline in on-site presence consistent with a shift to remote work; positive values indicate persistent physical attendance. 

The distribution of WPD differs markedly across clusters (Figure~\ref{fig:change_work_activity}). Central, office and institutional locations (\textit{Day shift}) experience the steepest decline on weekdays from 6AM to 8PM, with no variation on the weekends. Mixed-use locations (\textit{Mixed shift}) show a more moderate and variable response, with increased workplace presence during weekday night hours but a decrease during working hours. On the other hand, industrial locations experience increased daytime presence (\textit{Weekend shift}).  

These cluster-level differences in persistence are accompanied by systematic differences in industry composition, spatial location, and socio-economic characteristics (Figure~\ref{fig:radar_cluster}; see Methods for variable definitions). Central office locations, which empty most during the curfew, host the most productive firms but have a relatively lower number of companies that are less diverse in terms of their industries. These locations are also characterised by the lowest ratio of women in their daytime population. Industrial locations in the outskirts, which experience the most workplace dependence also attract the largest ratio of women and most uneven income distribution of workday presence. The central mixed-use spaces have the largest number of companies, with a high diversity in their industry, allowing for a good mix of visitors and workers from diverse economic backgrounds. These differences establish a clear empirical pattern: the locations most associated with knowledge-intensive, office-based work are also those whose daytime populations are most socio-economically stratified, and they are precisely the locations that emptied most during the curfew.

\begin{figure}[bh]
    \centering
    \includegraphics[width=\linewidth]{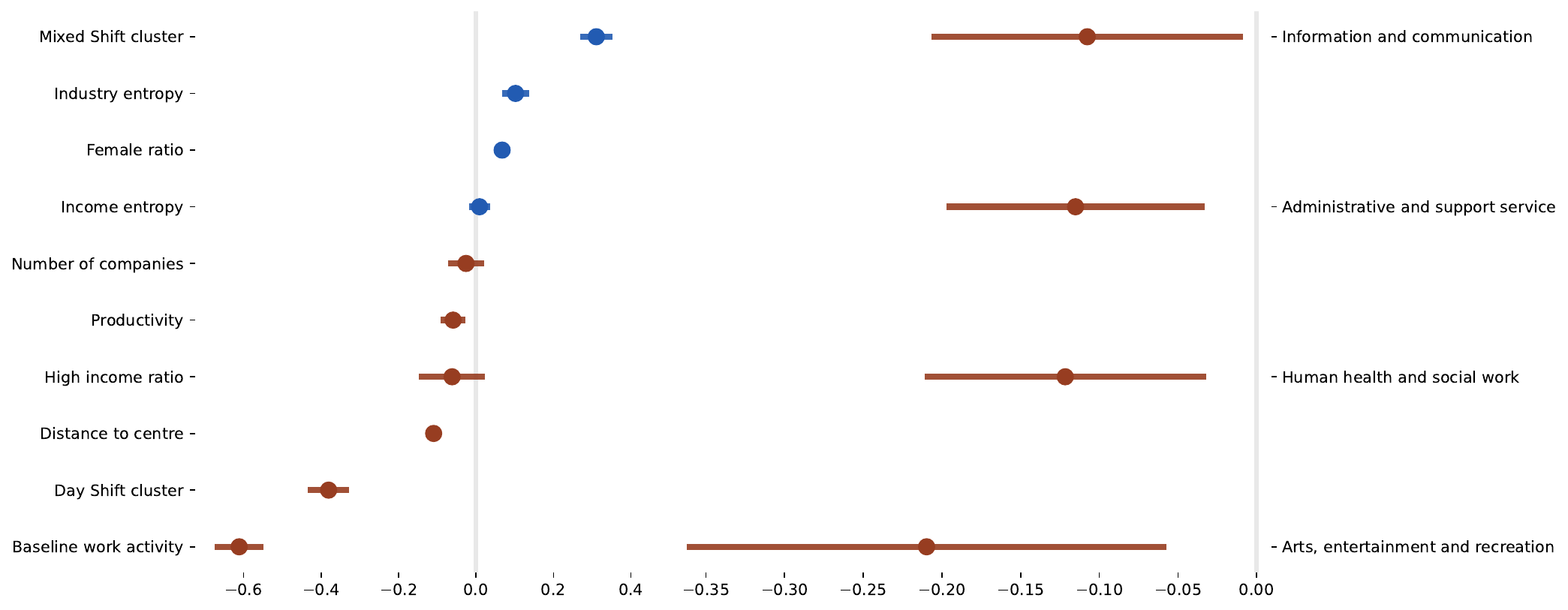}
    \caption{
        \textbf{Regression coefficients company, socio-economic and spatial characteristics. All variables were standardized. The error bars show 95\% confidence interval of the coefficient estimates. Full results are shown in the regression tables in SI Table S4 and S5.}
    }
    \label{fig:coef_plot}
\end{figure}

\subsection{Determinants of Workplace Dependence}\label{subsec:determinants}
To explore the contributions of industry composition, socio-economic characteristics, and spatial location to workplace persistence, we estimate an OLS regression using $\text{WPD}_i$, representing workplace dependence as the outcome. Prior to estimation, $\text{WPD}_i$ and all predictors are standardised to zero mean and unit standard deviation, so coefficients reflect effect sizes in comparable units.

\begin{equation}
\text{WPD}_i = \beta_0 + \boldsymbol{\beta_1} \mathbf{C}_i + 
\boldsymbol{\beta_2} \mathbf{S}_i + \boldsymbol{\beta_3} \mathbf{G}_i 
+ \beta_W W_i^{\text{open}} + \gamma_k + \varepsilon_i,
\label{eq:main_regression}
\end{equation}

where predictors are organised into three groups: company characteristics ($\boldsymbol{\beta_1} \mathbf{C}_i$), socio-economic characteristics of the daytime population ($\boldsymbol{\beta_2} \mathbf{S}_i$), and spatial characteristics ($\mathbf{S}_i + \boldsymbol{\beta_3} \mathbf{G}_i$) (see Methods for full model specification). Results are reported in Figure~\ref{fig:coef_plot} and in table format in the SI Table S4.

Company characteristics of a location are strong predictors of WPD as shown in Figure~\ref{fig:coef_plot}. Firm productivity is negatively associated with on-site presence ($\beta = -0.059$, $p < 0.01$): locations with more productive firms experienced steeper declines in attendance, suggesting that organisational capacity and digital infrastructure enabled more effective transitions to remote work. 
Industry composition is captured through sector fixed effects, with wholesale and retail trade (NACE sector G) as the reference category against which all other sectors are compared; the right-hand side of Figure~\ref{fig:coef_plot} reports the sectors with the strongest deviations from this baseline. Relative to wholesale and retail trade, Information and communication-dominated ($\beta = -0.108$, $p < 0.05$), as well as Administrative and support service heavy locations, show a strong decline in on-site attendance ($\beta = -0.115$, $p < 0.01$), consistent with the high remote-work potential of business service jobs \cite{dingel_how_2020, althoff_geography_2022}. At the same time, we also capture the reduction in WPD due to legally mandated closures and operational restrictions, including human health and social work ($\beta = -0.122$, $p < 0.01$) and the Arts, entertainment and recreation ($\beta = -0.210$, $p < 0.01$). This pattern aligns with the finding that primary and secondary industries remain more tied to physical presence than tertiary services \cite{hansen_remote_2023, yasenov_who_2020}.

\begin{figure}[h]
    \centering
    \begin{subfigure}[t]{\linewidth}
        \centering
        \includegraphics[width=\linewidth]{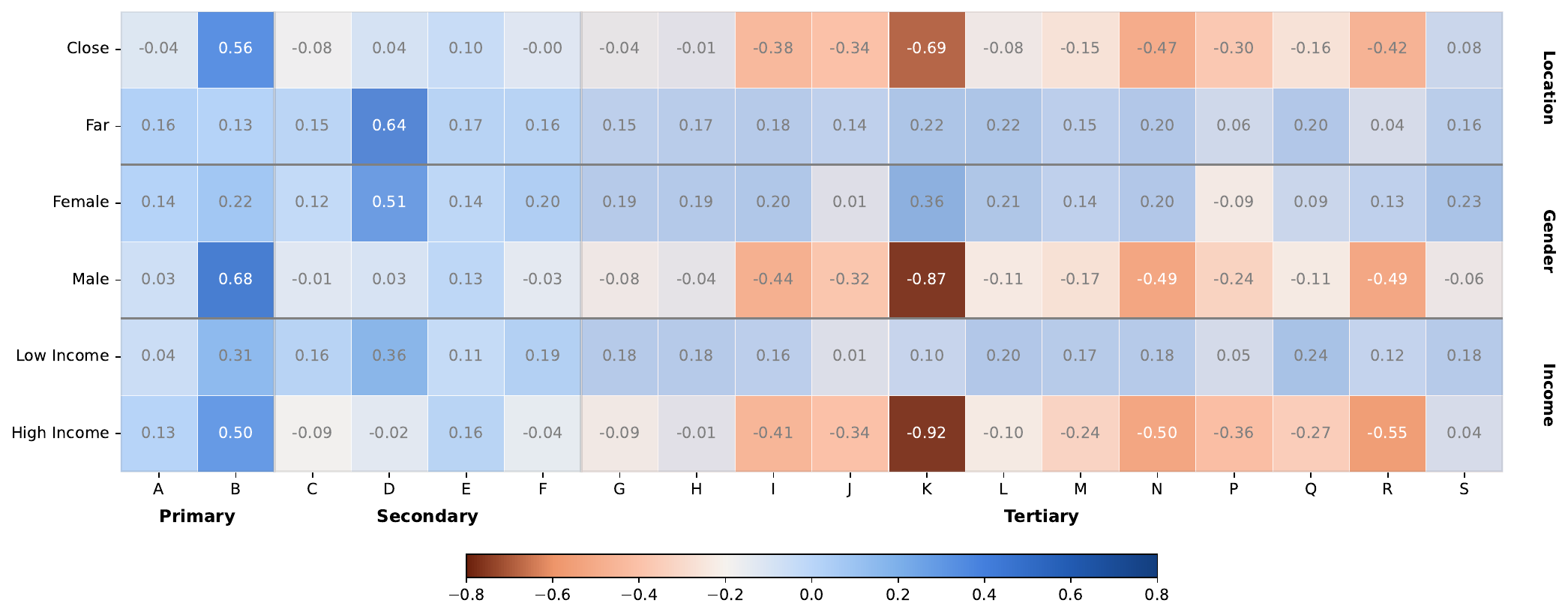}
        \caption{}
        \label{fig:heatmap_ind}
    \end{subfigure}
    \vspace{0.5em}
    \begin{subfigure}[t]{0.49\linewidth}
        \centering
        \includegraphics[width=\linewidth]{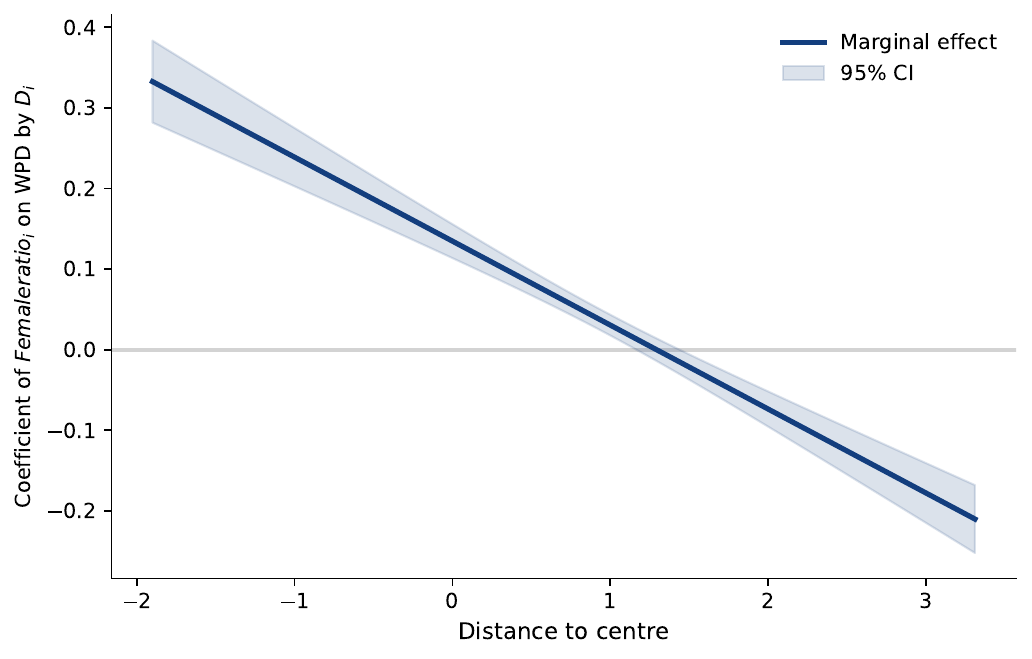}
        \caption{}
        \label{fig:fem_interplot}
    \end{subfigure}
    \hfill
    \begin{subfigure}[t]{0.49\linewidth}
        \centering
        \includegraphics[width=\linewidth]{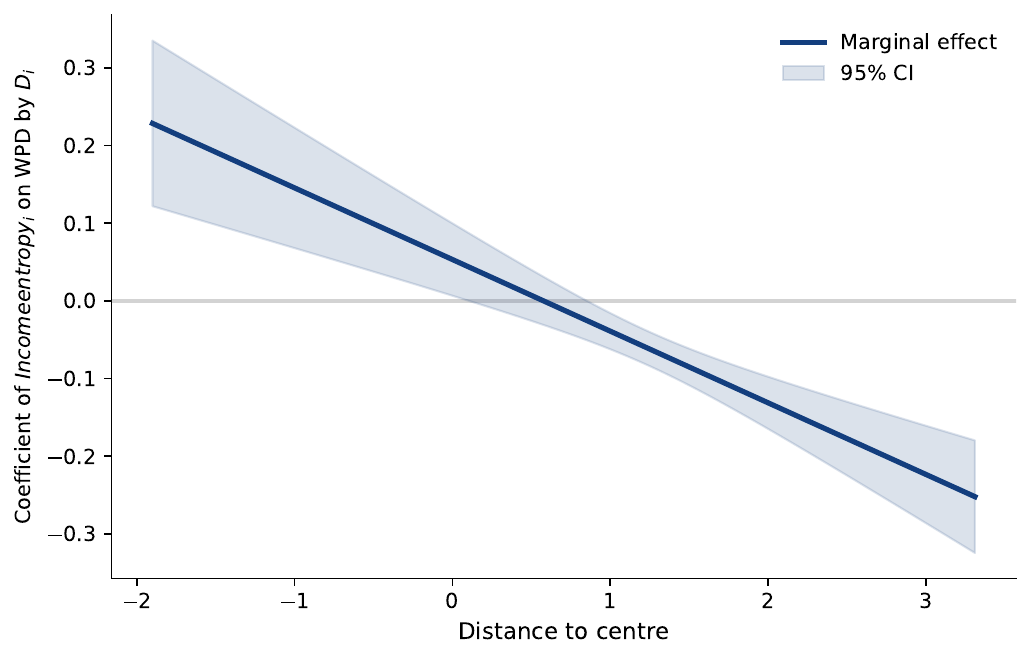}
        \caption{}
        \label{fig:income_interplot}
    \end{subfigure}
    \caption{
        \textbf{Understanding socio-economic variables}
        (\textbf{\subref{fig:heatmap_ind}}) Workplace dependence according to industry composition, where rowns represent location/gender/income strate, columns represent industry categories and colour represents WPD, where blue values mean higher average WPD. WPD is higher in primary and secondary industries, compared to tertiary industries, as well as female and low-income dominated locations.
        (\textbf{\subref{fig:fem_interplot}}) The effect of female ratio on workplace dependence declines with distance from the city centre
        (\textbf{\subref{fig:income_interplot}}) The effect of income diversity on workplace dependence declines with distance from the city centre
    }
    \label{fig:socio-econ}
\end{figure}

As Figure~\ref{fig:coef_plot} shows, the socio-economic composition of the daytime population during Opening time is a comparably strong predictor of WPD, and one with implications for social-mixing. Locations with a higher female presence show significantly higher WPD ($\beta = 0.067$, $p < 0.01$). The elevated WPD in areas with a female majority reflects occupational segregation, with women being concentrated in face-to-face roles such as retail, healthcare and care work rather than reflecting individual preferences for on-site work. 
Income composition tells a similar story: on average across the city, neither income entropy nor the ratio of high-income residents is significantly associated with WPD. However, both emerge as significant factors once their effect is allowed to vary with distance from the city centre, highlighting spatial differences. 
These effects indicate that locations characterised by mixed-income and female-majority daytime populations were the least able to shift to remote arrangements, sustaining on-site attendance even during strict COVID restrictions. 

Locations belonging to the major office and institutional hubs (\textit{Day shift}) show the steepest declines in WPD ($\beta = -0.381$, $p < 0.01$), confirming the descriptive pattern established in the previous subsection. Distance from the city centre is also negatively associated with workplace persistence ($\beta = -0.109$, $p < 0.01$): locations farther from the centre experienced smaller declines, or even gains, in WPD during the curfew. This result reflects the peripheral concentration of commerce and manufacturing activity, where remote work is structurally infeasible. The full model explains $R^2 = 0.354$ of the variance in $\text{workplace dependence}_i$. Moran's I on the OLS residuals confirms the presence of significant positive spatial autocorrelation ($I \approx$ 0.682, $p < 0.001$), indicating that neighbouring hexagons share residual variance not captured by the covariates. To assess whether this affects the substantive conclusions, we estimate a Spatial Error Model (SEM) in which the spatial structure of the error term is modelled explicitly (see Supplementary Table S6). The key coefficients remain substantively unchanged, confirming that the OLS estimates are robust to spatial dependence. Robustness checks, including stepwise regressions, and locations with only one company present are reported in the SI Tables S5 and S4.

The coefficients reported above, describe the effect for each variable across the whole city. However, these effects are not uniform across space. Figure~\ref{fig:heatmap_ind} shows that WPD varies systematically across industry categories and socio-economic strata: primary and secondary industries show stronger average WPD than tertiary, service industries, but within tertiary industries the finance and insurance column (NACE K) shows the steepest declines, consistent with their higher remote-work potential. Across most industry categories, locations with a higher proportion of women and lower-income workers show stronger WPD.

The association between female presence and WPD declines with distance from the city centre (Figure~\ref{fig:fem_interplot}). It is near the centre that female-majority locations show the strongest WPD. Further from the centre, female-majority locations are comparatively more able to shift to remote work, even when surrounded by businesses, where remote work is structurally infeasible. The relation is the same for income entropy (Figure~\ref{fig:income_interplot}); the marginal effect of income entropy is similarly positive near the centre and declines with distance. 
The pattern cannot be explained simply by industry composition: central locations concentrate industries with the highest potential for remote working, yet female presence remains a strong predictor of WPD there. The mechanism involves occupation: central, female-majority, income-diverse locations are disproportionately staffed by customer-facing workers whose roles support the urban core for a workforce that has itself become remote. This has consequences beyond individual workers. Central locations have historically functioned as sites of socio-economic mixing, where workers from different income groups encounter one another through shared workplaces and amenities. When remote-capable workers withdraw, the residual workforce becomes more socio-economically homogeneous, and the conditions for cross-class contact narrow rather than diversify. By contrast, peripheral female-majority and income-diverse locations retain a more heterogeneous mix of on-site workers and flexible remote workers living nearby. This provides the opportunity for localised social interaction across groups. This spatial moderation of socio-economic effects is the geographic basis of the 'service trap', which we discuss in more detail below: proximity to the centre compounds workplace dependence for those groups with the least capacity to work remotely.

\section{Discussion}\label{sec11}

Companies, public transport authorities, and urban planners around the world are working to understand and adapt to the new realities of work. This study contributes to that effort by examining how spatial location, industry composition, and socio-economic characteristics shaped workplace dependence in Budapest during  the shift to remote working.

We treat urban locations as integrated systems, where economic, social, and residential activities are interdependent. By combining company characteristics with population density data enriched by socio-economic variables, we move beyond identifying where workers maintain physical presence to understanding what this persistence reveals about urban functions and inequality.

Our findings confirm that industry composition is an important determinant of WPD, consistent with previous work showing that primary and secondary industries remain more tied to workplace presence than tertiary services \cite{dingel_how_2020, hansen_remote_2023, yasenov_who_2020}. However, industry alone does not explain the spatial patterning we observe. Locations dominated by female and lower-income daytime populations are consistently more likely to retain workers on-site, and this holds across industries. Similarly, locations farther from the city centre experience smaller declines in physical attendance. 

These results point to a systematic link between pre-existing socio-economic disadvantage and the inability to shift to remote work. Unlike prior approaches that infer remote-work capacity from home locations or job characteristics \cite{huang_staying_2022, dingel_how_2020}, our workplace-side measurement reveals that disadvantage is inscribed in the spatial fabric of where physical work persists.

The regularity of this spatial pattern suggests a systematic trend: workplace persistence did not merely reflect industry type, but was also mapped onto existing geographical disparities within the city. During working hours, women and lower-income workers tend to be located further from the centre, reflecting pre-existing inequalities. In the pre-pandemic setting, this peripheral positioning compounded disadvantage by reducing exposure to the higher-income, predominantly male populations concentrated in central workplaces. When the restrictions imposed by the onset of the pandemic took effect, central locations emptied while those on the periphery persisted. The workers who remained on-site were often the same groups who were already furthest from the opportunities concentrated in the centre.

The study has several limitations that should be noted before considering the broader implications. First, the temporal scope of the data constrains our measurement of change. Because data availability begins in June 2020, we cannot compare the curfew period to a true pre-pandemic baseline, and the measured difference between Opening and Curfew periods likely underestimates the full magnitude of pandemic-induced change. For sectors such as information and communications technology, where remote work transitions occurred earlier in the pandemic, the Opening period baseline may already reflect substantial remote work adoption, compressing the observable difference. Using September as a baseline, however, has the advantage of including already-remote workers rather than conflating remote work adoption with pandemic-related job losses in hospitality and retail. Second, Magyar Telekom uses subscription information to derive gender. This limits our analysis in two ways: we are only provided with binary sex categories (male/female), which does not cover the full gender spectrum, and the number of women in the dataset is likely under-represented due to the mismatch between the line owner and the actual phone user. 

Furthermore, Budapest represents a dense, historically monocentric European urban morphology. While our findings demonstrate a clear core-periphery gradient in workplace dependence, these dynamics may manifest differently in highly polycentric or auto-dependent cities, where the 'Donut Effect' might redistribute activity across multiple suburban nodes rather than a single periphery.
In polycentric cities, the same socio-economic filtering may still occur but distribute across multiple suburban nodes rather than along a single radial gradient, which would alter the interpretation of distance-based effects.

Despite these limitations, the consequences of this spatial shift are evident in urban form and function. Central office locations — usually the city's most socio-economically mixed spaces, where workers from different backgrounds encounter one another through commuting and shared amenities — saw the sharpest decline in daytime population. The departure of remote-capable workers leaves behind a centre populated disproportionately by lower-income and female face-to-face service workers, producing what we term a 'service trap' in which the city core's social composition narrows rather than diversifies. This double erosion — fewer knowledge workers cross-pollinating, and a residual workforce stratified rather than mixed — poses a long-term risk to the cross-class encounters that urban density is supposed to generate \cite{toth2021inequality}.

As the central core hollows out, the potential for social interaction appears to have migrated elsewhere. Peripheral locations with a female majority and income diversity remained active, sustaining local economies and creating conditions for new, more locally rooted forms of social interaction among on-site workers, remote workers living nearby, and unemployed residents with care responsibilities. Interestingly, a parallel dynamic emerged within the city centre itself: the residual attendance was highly concentrated in specific micro-areas populated by women and income-diverse groups, further evidencing the 'service trap'. Whether these emergent spaces generate comparable opportunities for social mixing across socio-economic groups remains to be seen, but they represent a meaningful shift in how urban proximity and exposure are organised.

Crucially, this disadvantage is not uniform across space. The association between female-majority and income-diverse daytime populations and workplace persistence is strongest near the city centre and weakens with distance (Fig.~\ref{fig:fem_interplot}, Fig.~\ref{fig:income_interplot}). Central workplace dependence therefore acts as a socio-economic marker of face-to-face service work, while peripheral workplace dependence is driven primarily by structural land-use — manufacturing, logistics, and industrial activity that requires physical presence regardless of who staffs it.

These considerations remain relevant even as many employers push for a return to the office (RTO). Post-pandemic RTO mandates have been uneven in their uptake; firms that are younger and led by younger CEOs continue to allow substantially higher rates of remote work \cite{aksoy_younger_2026}, and a large share of workers retain a strong preference for hybrid arrangements, with many willing to accept significant wage reductions to preserve them \cite{cullen_home_2025}. Evidence on the organisational consequences of remote work is mixed \cite{yang_effects_2021, brucks_virtual_2022, bernstein_how_2018, bloom_hybrid_2024}, but hybrid arrangements appear durable, with workers willing to trade significant wages to retain them \cite{cullen_home_2025} and younger firms continuing to permit them. Despite this ambiguity, it is estimated that working from home will not return to pre-pandemic patterns \cite{barrero_why_2021}, and people will continue to work in hybrid arrangements. Understanding how spatial working behaviour shapes urban exposure and economic activity therefore remains critically important.

Ultimately, these dynamics have implications for urban planning. 
A simple interpretation of the decline in the number of commuters might suggest a reduced demand for public transport and a decline in city centre economic activity. 
However, our findings in Budapest highlight that workplace activity in peripheral, industrial locations proved comparatively resistant to the shift toward remote work, sustaining demand for local amenities, retail, and services. 
Evidence suggests that home-based workers make more frequent local shopping trips and spend more on groceries at home \cite{baker_shopping_2026}, reinforcing the economic case for investing in accessible local centres rather than assuming that activity will inevitably reconcentrate in the city centre. 
These emerging peripheral hubs may also carry implications for social integration: as locations with a female majority and diverse incomes maintain daytime activity, they could support new, localized encounters -- whether this translates into social mixing remains an open question for future work. 
Investment in these locations through the creation of amenities and transit may contribute towards the development of mixed-use centres.
These shifts are consistent with a broader move toward a more multicentric city, requiring new strategies to connect and support these redistributed hubs of activity.

\section{Methods}\label{sec11}
\subsection{Temporal population density data}\label{subsec2}
Our geolocated data comes from Magyar Telekom, a leading Hungarian telecommunications service provider. It provides aggregated population density data at a 100x100 m hexagonal level, based on device traffic at one-hourly intervals, from 02/06/2020 to 31/05/2021. This population density data is derived from the raw geolocated data available to the company and is based on telecommunications events occurring on 2G, 3G and 4G networks on mobile phones and tablets only (e.g. phone calls, text messages, internet usage and movement between cell tower areas). These events are used to locate the area via triangulation. Devices are linked to demographic data derived from anonymised subscriber databases. The data was provided in accordance with the GDPR data protection regulation through a secure SFTP server.

We are only considering data from 1 September to 31 October and 11 November to 23 December 2020. We are considering this period because the Covid emergency period ended on 17 June and new restrictions were introduced on 10 November when teaching moved online and an 8PM curfew was introduced. The 1st of September also marks the start of the school year, enabling us to select a relatively normal period. Since 24 December is the start of the holiday period, we also excluded the winter holidays to reduce noise relating to work location and general movement. 

\subsection{Spatial and temporal smoothing}\label{subsec3}
The triangulation approach provides a reasonable, albeit inexact, location of people. To correct for this, we carry out spatial smoothing based on a moving average. We calculate the average number of people in a hexagon and its six neighbours, moving through all the hexagons in Budapest. This ensures a more even distribution.

Secondly, in order to comply with GDPR regulations, Hungarian Telekom does not provide information for locations with fewer than 10 activities. For hexagons with a relatively low number of activities, this may result in significant fluctuations in presence, with values jumping from 10 to 0, from one hour to the next. To navigate such unpredictability, we use Gaussian smoothing. 

\subsection{Creating Opening and Curfew times}\label{subsec3}
After smoothing the relevant timeframes spatially and temporally, we create an average 48-hour period, where the first 24 hours represent an average weekday, while the next 24 hours represent an average weekend. This ensures that we capture all workers, including those undertaking hybrid work, and provides an overview of the population at a given location and hour for a typical weekday, and weekend. We have chosen a 48-hour period because we differentiate between weekday and weekend activity, and we expect different working and activity trends based on the day of the week. 

\subsection{Clustering}\label{subsec3}
The company identifies home and work locations of individuals, and shares the number of people at these locations for each hour. Using this workplace presence information, we identify 3 different work patterns. To do this, we do a KMeans clustering on the temporal working activity of the logarithm of the number of workers at a location during the 48 hour opening period. We validate the results both by plotting the average temporal work activity and by connecting the clustering data to available company information and we see qualitative differences.

\subsection{Variable definitions}\label{subsec3}

We identify 4 types of key variables for understanding what impacts the change in WPD. First, we focus on company variables, and geolocate companies with at least 3 employees from OPTEN to h3 hexagon levels. We count the number of companies in each location. Then, we focus on understanding their industry characteristics. We construct an indicator of industry dominance at the location level. For each location, let $E_i$ denote total employment in industry $i$, and let $K$ be the set of industries present. The dominant industry is defined as the industry with the highest level of employment: 

\begin{equation}
i^{*} = \arg\max_{i \in K} E_i
\end{equation} 

In addition to industry dominance, we measure the industry diversity of firms within a location using Shannon entropy. 
For each location, entropy is defined as:

\begin{equation}
H = - \sum_{i=1}^{K} p_i \log(p_i)
\end{equation}

where $p_i$ denotes the proportion of firms belonging to industry $i$ in that location, and $K$ is the total number of distinct industries present. High values mean that the location is diversified, companies spread more evenly across industries, and a low entropy means the location is specialised; dominated by few industries. 
We measure average productivity at the location level as the mean of firm-level productivity. For each firm $j$, productivity is defined as revenue per worker:

\begin{equation}
\text{prod}_j = \frac{\text{Revenue}_j}{\text{Employment}_j}
\end{equation}

Average productivity within a location containing $N$ firms is then computed as:
\begin{equation}
\overline{prod} = \frac{1}{N} \sum_{j=1}^{N} \pi_j
\end{equation}

This measure reflects the average productivity of firms operating in a given location, treating each firm equally, regardless of its size.

Secondly, we identify the presence of people according to their activity and socio-economic background. For each location, we count:
a) the number of people,
b) the number of people at their place of work,
c) the ratio of women,
d) the ratio of high income individuals, and
e) income entropy.
These data are derived from the Telekom population density dataset, which includes identification of gender, income level, and work and home location. Income entropy is calculated as the Shannon entropy of people belonging to three income categories (low-, middle-, and high-income) within a specific geographical location. High values indicate income diversity, while low values highlight income inequality. Our approach follows recent frameworks that treat income segregation as a dynamic, time-varying property of urban space—driven by daily activity—rather than a static residential characteristic \cite{rossi2025time}.

Finally, the distance from the centre is calculated by measuring the geodesic distance (in metres) between the centroid of each H3 hexagon and Deák Ferenc Square.

\subsection{Statistical models}\label{subsec3}
To understand the company, socio-economic and spatial heterogeneity in the decreased workplace presence in 2020 curfew time, compared to baseline opening time, we model $\text{WPD}_i$ using an 
ordinary least squares (OLS) regression with HC3 heteroskedasticity-robust 
standard errors. We include variables describing the industry composition 
of a location using full industry fixed effects, with wholesale and retail 
trade (NACE sector G) as the reference category, alongside average 
productivity and number of companies, weighted by employee count where 
locations contain more than one firm. For socio-economic variables, we 
include income entropy, the ratio of women, and the ratio of high-income 
individuals present during opening time. We additionally include the 
spatial cluster membership of each location (\textit{Day Shift} and \textit{Mixed Shift}, 
with \textit{Weekend Shift} as reference), and control for baseline work activity 
during opening time and distance from the city centre. All predictors 
are standardised prior to estimation. We measure all variables at 10AM 
on the average weekday, as most workers are recorded at this time, on \ref{eq:main_regression}, where $\text{WPD}_i$ represents workplace dependence at 
location $i$, measured as the difference between the average number 
of workers during curfew time (November--December 2020) and opening 
time (September--October 2020) at 10AM on a workday. $W_i^{\text{open}}$ 
denotes the spatially smoothed baseline work activity during opening 
time, and $\gamma_k$ denotes industry fixed effects for NACE sector 
$k$, with sector G (wholesale and retail trade) as the reference 
category. The explanatory variables are grouped into three categories:

\paragraph{Company characteristics ($\boldsymbol{\beta_1} \mathbf{C}_i$):}
\begin{align*}
\mathbf{C}_i = \{ &\text{Number of companies}_i,\ 
                   \text{Productivity}_i,\ 
                   \gamma_k \}
\end{align*}
\noindent where $\text{Number of companies}_i$ indicates the total 
number of firms registered at the location, $\text{Productivity}_i$ 
measures average company-level productivity per employee, and $\gamma_k$ 
captures industry fixed effects across all NACE sectors (A--S), with 
sector G as the reference category.

\paragraph{Socio-economic characteristics ($\boldsymbol{\beta_2} \mathbf{S}_i$):}
\begin{align*}
\mathbf{S}_i = \{ &\text{Income entropy}_i,\ 
                   \text{Female ratio}_i,\ 
                   \text{High income ratio}_i \}
\end{align*}
\noindent where $\text{Income entropy}_i$ measures the Shannon entropy 
of income groups present at the location during opening time, capturing 
the diversity of the daytime income distribution; $\text{Female ratio}_i$ 
is the proportion of women present during opening time; and 
$\text{High income ratio}_i$ is the proportion of high-income 
individuals present during opening time.

\paragraph{Spatial characteristics ($\boldsymbol{\beta_3} \mathbf{G}_i$):}
\begin{align*}
\mathbf{G}_i = \{ &\text{Distance}_i,\ 
                   \text{Day Shift}_i,\ 
                   \text{Mixed Shift}_i \}
\end{align*}
\noindent where $\text{Distance}_i$ is the Euclidean distance to 
De\'{a}k Ferenc t\'{e}r in the city centre; $\text{Day Shift}_i$ and 
$\text{Mixed Shift}_i$ are binary indicators for membership in the 
Day Shift and Mixed Shift spatial work clusters respectively, derived 
from the temporal clustering described in the previous section, with 
Weekend Shift as the reference category.

To examine whether the relationship between socio-economic 
characteristics and WPD varies with distance from the city centre, 
we estimate two interaction models extending 
Equation~\ref{eq:main_regression}:

\begin{equation}
\text{WPD}_i = \beta_0 + \boldsymbol{\beta_1} \mathbf{C}_i + 
\boldsymbol{\beta_2} \mathbf{S}_i + \boldsymbol{\beta_3} \mathbf{G}_i 
+ \beta_W W_i^{\text{open}} + \beta_{F} \cdot \text{Female ratio}_i \cdot 
\text{Distance}_i + \gamma_k + \varepsilon_i
\label{eq:interaction_F}
\end{equation}

\begin{equation}
\text{WPD}_i = \beta_0 + \boldsymbol{\beta_1} \mathbf{C}_i + 
\boldsymbol{\beta_2} \mathbf{S}_i + \boldsymbol{\beta_3} \mathbf{G}_i 
+ \beta_W W_i^{\text{open}} + \beta_{H} \cdot \text{Income entropy}_i \cdot 
\text{Distance}_i + \gamma_k + \varepsilon_i
\label{eq:interaction_H}
\end{equation}

\noindent Equations~\ref{eq:interaction_F} and~\ref{eq:interaction_H} 
are estimated separately, each extending the baseline 
model (Equation~\ref{eq:main_regression}) with a single interaction 
term testing whether the association between female ratio and WPD, 
and income entropy and WPD respectively, is moderated by distance 
from the city centre.

\backmatter

\subsection*{Acknowledgements}
The authors are grateful for Bence Kovács's geolocation of company data, comments and suggestions of Bence Pálfi, Nandini Iyer, Esteban Moro, Gergő Tóth and participants of the IC2S2 and Geography of Innovation conferences. B.L. acknowledges financial help received from the MTA Lendület Award. The authors acknowledge financial help from NRDI Office of Hungary through the Driving Urban Transition European Partnership project COLINE (Complex Links of Neighbourhoods F-DUT-2023-0037). R.D.C. acknowledges support from the Lagrange Project of the ISI Foundation funded by CRT Foundation. 

\subsection*{Ethics declarations}
The authors declare no competing interests.

\subsection*{Data availability}
Data sufficient to reproduce all results in this paper will be made available upon request.

\subsection*{Code availability}
The analysis was conducted using Python. The code to reproduce the main results in the figures from the aggregated data is publicly available on GitHub: \url{https://github.com/zzsofi/Workplace_dependence_in_urban_economies}.

\subsection*{Author contribution}
Z.Z., R.D.C., and B.L. designed the research. Z.Z. performed the analysis.  Z.Z, B.L and R.D.C. wrote the paper. R.D.C. and B.L. supervised the project. All authors discussed the results and contributed to the final manuscript.

\bibliography{sn-bibliography}


\begin{thebibliography}{45}
\ifx \bisbn   \undefined \def \bisbn  #1{ISBN #1}\fi
\ifx \binits  \undefined \def \binits#1{#1}\fi
\ifx \bauthor  \undefined \def \bauthor#1{#1}\fi
\ifx \batitle  \undefined \def \batitle#1{#1}\fi
\ifx \bjtitle  \undefined \def \bjtitle#1{#1}\fi
\ifx \bvolume  \undefined \def \bvolume#1{\textbf{#1}}\fi
\ifx \byear  \undefined \def \byear#1{#1}\fi
\ifx \bissue  \undefined \def \bissue#1{#1}\fi
\ifx \bfpage  \undefined \def \bfpage#1{#1}\fi
\ifx \blpage  \undefined \def \blpage #1{#1}\fi
\ifx \burl  \undefined \def \burl#1{\textsf{#1}}\fi
\ifx \doiurl  \undefined \def \doiurl#1{\url{https://doi.org/#1}}\fi
\ifx \betal  \undefined \def \betal{\textit{et al.}}\fi
\ifx \binstitute  \undefined \def \binstitute#1{#1}\fi
\ifx \binstitutionaled  \undefined \def \binstitutionaled#1{#1}\fi
\ifx \bctitle  \undefined \def \bctitle#1{#1}\fi
\ifx \beditor  \undefined \def \beditor#1{#1}\fi
\ifx \bpublisher  \undefined \def \bpublisher#1{#1}\fi
\ifx \bbtitle  \undefined \def \bbtitle#1{#1}\fi
\ifx \bedition  \undefined \def \bedition#1{#1}\fi
\ifx \bseriesno  \undefined \def \bseriesno#1{#1}\fi
\ifx \blocation  \undefined \def \blocation#1{#1}\fi
\ifx \bsertitle  \undefined \def \bsertitle#1{#1}\fi
\ifx \bsnm \undefined \def \bsnm#1{#1}\fi
\ifx \bsuffix \undefined \def \bsuffix#1{#1}\fi
\ifx \bparticle \undefined \def \bparticle#1{#1}\fi
\ifx \barticle \undefined \def \barticle#1{#1}\fi
\bibcommenthead
\ifx \bconfdate \undefined \def \bconfdate #1{#1}\fi
\ifx \botherref \undefined \def \botherref #1{#1}\fi
\ifx \url \undefined \def \url#1{\textsf{#1}}\fi
\ifx \bchapter \undefined \def \bchapter#1{#1}\fi
\ifx \bbook \undefined \def \bbook#1{#1}\fi
\ifx \bcomment \undefined \def \bcomment#1{#1}\fi
\ifx \oauthor \undefined \def \oauthor#1{#1}\fi
\ifx \citeauthoryear \undefined \def \citeauthoryear#1{#1}\fi
\ifx \endbibitem  \undefined \def \endbibitem {}\fi
\ifx \bconflocation  \undefined \def \bconflocation#1{#1}\fi
\ifx \arxivurl  \undefined \def \arxivurl#1{\textsf{#1}}\fi
\csname PreBibitemsHook\endcsname

\bibitem[\protect\citeauthoryear{Barrero et~al.}{2023}]{barrero_evolution_2023}
\begin{barticle}
\bauthor{\bsnm{Barrero}, \binits{J.M.}},
\bauthor{\bsnm{Bloom}, \binits{N.}},
\bauthor{\bsnm{Davis}, \binits{S.J.}}:
\batitle{The {Evolution} of {Work} from {Home}}.
\bjtitle{Journal of Economic Perspectives}
\bvolume{37}(\bissue{4}),
\bfpage{23}--\blpage{49}
(\byear{2023})
\doiurl{10.1257/jep.37.4.23}
\end{barticle}
\endbibitem

\bibitem[\protect\citeauthoryear{Bick et~al.}{2023}]{bick_work_2023}
\begin{barticle}
\bauthor{\bsnm{Bick}, \binits{A.}},
\bauthor{\bsnm{Blandin}, \binits{A.}},
\bauthor{\bsnm{Mertens}, \binits{K.}}:
\batitle{Work from {Home} before and after the {COVID}-19 {Outbreak}}.
\bjtitle{American Economic Journal: Macroeconomics}
\bvolume{15}(\bissue{4}),
\bfpage{1}--\blpage{39}
(\byear{2023})
\doiurl{10.1257/mac.20210061}
\end{barticle}
\endbibitem

\bibitem[\protect\citeauthoryear{Barrero et~al.}{2021}]{barrero_why_2021}
\begin{botherref}
\oauthor{\bsnm{Barrero}, \binits{J.M.}},
\oauthor{\bsnm{Bloom}, \binits{N.}},
\oauthor{\bsnm{Davis}, \binits{S.}}:
Why {Working} from {Home} {Will} {Stick}.
Technical Report w28731,
National Bureau of Economic Research,
Cambridge, MA
(April 2021).
\doiurl{10.3386/w28731} .
\url{http://www.nber.org/papers/w28731.pdf}
\end{botherref}
\endbibitem

\bibitem[\protect\citeauthoryear{Santana et~al.}{2023}]{santana2023covid}
\begin{barticle}
\bauthor{\bsnm{Santana}, \binits{C.}},
\bauthor{\bsnm{Botta}, \binits{F.}},
\bauthor{\bsnm{Barbosa}, \binits{H.}},
\bauthor{\bsnm{Privitera}, \binits{F.}},
\bauthor{\bsnm{Menezes}, \binits{R.}},
\bauthor{\bsnm{Di~Clemente}, \binits{R.}}:
\batitle{Covid-19 is linked to changes in the time--space dimension of human mobility}.
\bjtitle{Nature Human Behaviour}
\bvolume{7}(\bissue{10}),
\bfpage{1729}--\blpage{1739}
(\byear{2023})
\end{barticle}
\endbibitem

\bibitem[\protect\citeauthoryear{Dingel and Neiman}{2020}]{dingel_how_2020}
\begin{barticle}
\bauthor{\bsnm{Dingel}, \binits{J.I.}},
\bauthor{\bsnm{Neiman}, \binits{B.}}:
\batitle{How many jobs can be done at home?}
\bjtitle{Journal of Public Economics}
\bvolume{189},
\bfpage{104235}
(\byear{2020})
\doiurl{10.1016/j.jpubeco.2020.104235}
\end{barticle}
\endbibitem

\bibitem[\protect\citeauthoryear{Koren and Pet{\H{o}}}{2020}]{koren2020business}
\begin{barticle}
\bauthor{\bsnm{Koren}, \binits{M.}},
\bauthor{\bsnm{Pet{\H{o}}}, \binits{R.}}:
\batitle{Business disruptions from social distancing}.
\bjtitle{Plos one}
\bvolume{15}(\bissue{9}),
\bfpage{0239113}
(\byear{2020})
\end{barticle}
\endbibitem

\bibitem[\protect\citeauthoryear{Adams-Prassl et~al.}{2022}]{adams2022work}
\begin{barticle}
\bauthor{\bsnm{Adams-Prassl}, \binits{A.}},
\bauthor{\bsnm{Boneva}, \binits{T.}},
\bauthor{\bsnm{Golin}, \binits{M.}},
\bauthor{\bsnm{Rauh}, \binits{C.}}:
\batitle{Work that can be done from home: Evidence on variation within and across occupations and industries}.
\bjtitle{Labour Economics}
\bvolume{74},
\bfpage{102083}
(\byear{2022})
\end{barticle}
\endbibitem

\bibitem[\protect\citeauthoryear{Minkus et~al.}{2022}]{minkus2022significance}
\begin{barticle}
\bauthor{\bsnm{Minkus}, \binits{L.}},
\bauthor{\bsnm{Groepler}, \binits{N.}},
\bauthor{\bsnm{Drobni{\v{c}}}, \binits{S.}}:
\batitle{The significance of occupations, family responsibilities, and gender for working from home: Lessons from covid-19}.
\bjtitle{PLoS One}
\bvolume{17}(\bissue{6}),
\bfpage{0266393}
(\byear{2022})
\end{barticle}
\endbibitem

\bibitem[\protect\citeauthoryear{Huang et~al.}{2022}]{huang_staying_2022}
\begin{barticle}
\bauthor{\bsnm{Huang}, \binits{X.}},
\bauthor{\bsnm{Lu}, \binits{J.}},
\bauthor{\bsnm{Gao}, \binits{S.}},
\bauthor{\bsnm{Wang}, \binits{S.}},
\bauthor{\bsnm{Liu}, \binits{Z.}},
\bauthor{\bsnm{Wei}, \binits{H.}}:
\batitle{Staying at {Home} {Is} a {Privilege}: {Evidence} from {Fine}-{Grained} {Mobile} {Phone} {Location} {Data} in the {United} {States} during the {COVID}-19 {Pandemic}}.
\bjtitle{Annals of the American Association of Geographers}
\bvolume{112}(\bissue{1}),
\bfpage{286}--\blpage{305}
(\byear{2022})
\doiurl{10.1080/24694452.2021.1904819}
\end{barticle}
\endbibitem

\bibitem[\protect\citeauthoryear{Lunde et~al.}{2022}]{lunde2022relationship}
\begin{barticle}
\bauthor{\bsnm{Lunde}, \binits{L.-K.}},
\bauthor{\bsnm{Fl{\o}vik}, \binits{L.}},
\bauthor{\bsnm{Christensen}, \binits{J.O.}},
\bauthor{\bsnm{Johannessen}, \binits{H.A.}},
\bauthor{\bsnm{Finne}, \binits{L.B.}},
\bauthor{\bsnm{J{\o}rgensen}, \binits{I.L.}},
\bauthor{\bsnm{Mohr}, \binits{B.}},
\bauthor{\bsnm{Vleeshouwers}, \binits{J.}}:
\batitle{The relationship between telework from home and employee health: a systematic review}.
\bjtitle{BMC public health}
\bvolume{22}(\bissue{1}),
\bfpage{47}
(\byear{2022})
\end{barticle}
\endbibitem

\bibitem[\protect\citeauthoryear{Gozzi et~al.}{2021}]{gozzi_estimating_2021}
\begin{barticle}
\bauthor{\bsnm{Gozzi}, \binits{N.}},
\bauthor{\bsnm{Tizzoni}, \binits{M.}},
\bauthor{\bsnm{Chinazzi}, \binits{M.}},
\bauthor{\bsnm{Ferres}, \binits{L.}},
\bauthor{\bsnm{Vespignani}, \binits{A.}},
\bauthor{\bsnm{Perra}, \binits{N.}}:
\batitle{Estimating the effect of social inequalities on the mitigation of {COVID}-19 across communities in {Santiago} de {Chile}}.
\bjtitle{Nature Communications}
\bvolume{12}(\bissue{1}),
\bfpage{2429}
(\byear{2021})
\doiurl{10.1038/s41467-021-22601-6}
\end{barticle}
\endbibitem

\bibitem[\protect\citeauthoryear{Aksoy et~al.}{2022}]{aksoy_working_2022}
\begin{barticle}
\bauthor{\bsnm{Aksoy}, \binits{C.G.}},
\bauthor{\bsnm{Barrero}, \binits{J.M.}},
\bauthor{\bsnm{Bloom}, \binits{N.}},
\bauthor{\bsnm{Davis}, \binits{S.J.}},
\bauthor{\bsnm{Dolls}, \binits{M.}},
\bauthor{\bsnm{Zarate}, \binits{P.}}:
\batitle{Working from {Home} {Around} the {World}}.
\bjtitle{Brookings Papers on Economic Activity}
\bvolume{2022}(\bissue{2}),
\bfpage{281}--\blpage{360}
(\byear{2022})
\doiurl{10.1353/eca.2022.a901274}
\end{barticle}
\endbibitem

\bibitem[\protect\citeauthoryear{Althoff et~al.}{2022}]{althoff_geography_2022}
\begin{barticle}
\bauthor{\bsnm{Althoff}, \binits{L.}},
\bauthor{\bsnm{Eckert}, \binits{F.}},
\bauthor{\bsnm{Ganapati}, \binits{S.}},
\bauthor{\bsnm{Walsh}, \binits{C.}}:
\batitle{The {Geography} of {Remote} {Work}}.
\bjtitle{Regional Science and Urban Economics}
\bvolume{93},
\bfpage{103770}
(\byear{2022})
\doiurl{10.1016/j.regsciurbeco.2022.103770}
\end{barticle}
\endbibitem

\bibitem[\protect\citeauthoryear{Ramani et~al.}{2024}]{ramani_how_2024}
\begin{barticle}
\bauthor{\bsnm{Ramani}, \binits{A.}},
\bauthor{\bsnm{Alcedo}, \binits{J.}},
\bauthor{\bsnm{Bloom}, \binits{N.}}:
\batitle{How working from home reshapes cities}.
\bjtitle{Proceedings of the National Academy of Sciences}
\bvolume{121}(\bissue{45}),
\bfpage{2408930121}
(\byear{2024})
\doiurl{10.1073/pnas.2408930121}
\end{barticle}
\endbibitem

\bibitem[\protect\citeauthoryear{Gupta et~al.}{2026}]{gupta2026work}
\begin{barticle}
\bauthor{\bsnm{Gupta}, \binits{A.}},
\bauthor{\bsnm{Mittal}, \binits{V.}},
\bauthor{\bsnm{Van~Nieuwerburgh}, \binits{S.}}:
\batitle{Work from home and the office real estate apocalypse}.
\bjtitle{American Economic Review}
\bvolume{116}(\bissue{2}),
\bfpage{674}--\blpage{709}
(\byear{2026})
\end{barticle}
\endbibitem

\bibitem[\protect\citeauthoryear{Yabe et~al.}{2024}]{yabe_behaviour-based_2024}
\begin{barticle}
\bauthor{\bsnm{Yabe}, \binits{T.}},
\bauthor{\bsnm{García Bulle~Bueno}, \binits{B.}},
\bauthor{\bsnm{Frank}, \binits{M.R.}},
\bauthor{\bsnm{Pentland}, \binits{A.}},
\bauthor{\bsnm{Moro}, \binits{E.}}:
\batitle{Behaviour-based dependency networks between places shape urban economic resilience}.
\bjtitle{Nature Human Behaviour}
(\byear{2024})
\doiurl{10.1038/s41562-024-02072-7} .
\bcomment{Publisher: Springer Science and Business Media LLC}
\end{barticle}
\endbibitem

\bibitem[\protect\citeauthoryear{Yabe et~al.}{2023}]{yabe_behavioral_2023}
\begin{botherref}
\oauthor{\bsnm{Yabe}, \binits{T.}},
\oauthor{\bsnm{Bueno}, \binits{B.G.B.}},
\oauthor{\bsnm{Dong}, \binits{X.}},
\oauthor{\bsnm{Pentland}, \binits{A.}},
\oauthor{\bsnm{Moro}, \binits{E.}}:
Behavioral changes during the {COVID}-19 pandemic decreased income diversity of urban encounters.
Nature Communications
\textbf{14}(1)
(2023)
\doiurl{10.1038/s41467-023-37913-y} .
Publisher: Springer Science and Business Media LLC
\end{botherref}
\endbibitem

\bibitem[\protect\citeauthoryear{Alonso}{1964}]{alonso1964location}
\begin{botherref}
\oauthor{\bsnm{Alonso}, \binits{W.}}:
Location theory.
Cambridge MA: MIT Press
(1964)
\end{botherref}
\endbibitem

\bibitem[\protect\citeauthoryear{Moretti}{2012}]{moretti2012new}
\begin{bbook}
\bauthor{\bsnm{Moretti}, \binits{E.}}:
\bbtitle{The New Geography of Jobs}.
\bpublisher{Houghton Mifflin Harcourt},
\blocation{Boston}
(\byear{2012})
\end{bbook}
\endbibitem

\bibitem[\protect\citeauthoryear{Glaeser and Mar{\'e}}{2001}]{glaeser2001cities}
\begin{barticle}
\bauthor{\bsnm{Glaeser}, \binits{E.L.}},
\bauthor{\bsnm{Mar{\'e}}, \binits{D.C.}}:
\batitle{Cities and skills}.
\bjtitle{Journal of labor economics}
\bvolume{19}(\bissue{2}),
\bfpage{316}--\blpage{342}
(\byear{2001})
\end{barticle}
\endbibitem

\bibitem[\protect\citeauthoryear{Zhong et~al.}{2017}]{zhong2017revealing}
\begin{barticle}
\bauthor{\bsnm{Zhong}, \binits{C.}},
\bauthor{\bsnm{Schl{\"a}pfer}, \binits{M.}},
\bauthor{\bsnm{M{\"u}ller~Arisona}, \binits{S.}},
\bauthor{\bsnm{Batty}, \binits{M.}},
\bauthor{\bsnm{Ratti}, \binits{C.}},
\bauthor{\bsnm{Schmitt}, \binits{G.}}:
\batitle{Revealing centrality in the spatial structure of cities from human activity patterns}.
\bjtitle{Urban Studies}
\bvolume{54}(\bissue{2}),
\bfpage{437}--\blpage{455}
(\byear{2017})
\end{barticle}
\endbibitem

\bibitem[\protect\citeauthoryear{Moro et~al.}{2021}]{moro_mobility_2021}
\begin{botherref}
\oauthor{\bsnm{Moro}, \binits{E.}},
\oauthor{\bsnm{Calacci}, \binits{D.}},
\oauthor{\bsnm{Dong}, \binits{X.}},
\oauthor{\bsnm{Pentland}, \binits{A.}}:
Mobility patterns are associated with experienced income segregation in large {US} cities.
Nature Communications
\textbf{12}(1)
(2021)
\doiurl{10.1038/s41467-021-24899-8} .
Publisher: Springer Science and Business Media LLC
\end{botherref}
\endbibitem

\bibitem[\protect\citeauthoryear{Juh{\'a}sz et~al.}{2023}]{juhasz2023amenity}
\begin{barticle}
\bauthor{\bsnm{Juh{\'a}sz}, \binits{S.}},
\bauthor{\bsnm{Pint{\'e}r}, \binits{G.}},
\bauthor{\bsnm{Kov{\'a}cs}, \binits{{\'A}.J.}},
\bauthor{\bsnm{Borza}, \binits{E.}},
\bauthor{\bsnm{M{\'o}nus}, \binits{G.}},
\bauthor{\bsnm{L{\H{o}}rincz}, \binits{L.}},
\bauthor{\bsnm{Lengyel}, \binits{B.}}:
\batitle{Amenity complexity and urban locations of socio-economic mixing}.
\bjtitle{EPJ Data Science}
\bvolume{12}(\bissue{1}),
\bfpage{34}
(\byear{2023})
\end{barticle}
\endbibitem

\bibitem[\protect\citeauthoryear{Jacobs}{1972}]{jacobs_death_1972}
\begin{bbook}
\bauthor{\bsnm{Jacobs}, \binits{J.}}:
\bbtitle{The Death and Life of Great {American} Cities}.
\bsertitle{Pelican books}.
\bpublisher{Penguin},
\blocation{Harmondsworth}
(\byear{1972})
\end{bbook}
\endbibitem

\bibitem[\protect\citeauthoryear{Brueckner et~al.}{1999}]{brueckner1999central}
\begin{barticle}
\bauthor{\bsnm{Brueckner}, \binits{J.K.}},
\bauthor{\bsnm{Thisse}, \binits{J.-F.}},
\bauthor{\bsnm{Zenou}, \binits{Y.}}:
\batitle{Why is central paris rich and downtown detroit poor?: An amenity-based theory}.
\bjtitle{European economic review}
\bvolume{43}(\bissue{1}),
\bfpage{91}--\blpage{107}
(\byear{1999})
\end{barticle}
\endbibitem

\bibitem[\protect\citeauthoryear{Zhang and Pryce}{2020}]{zhang2020dynamics}
\begin{barticle}
\bauthor{\bsnm{Zhang}, \binits{M.L.}},
\bauthor{\bsnm{Pryce}, \binits{G.}}:
\batitle{The dynamics of poverty, employment and access to amenities in polycentric cities: Measuring the decentralisation of poverty and its impacts in england and wales}.
\bjtitle{Urban Studies}
\bvolume{57}(\bissue{10}),
\bfpage{2015}--\blpage{2030}
(\byear{2020})
\end{barticle}
\endbibitem

\bibitem[\protect\citeauthoryear{White}{1986}]{white1986sex}
\begin{barticle}
\bauthor{\bsnm{White}, \binits{M.J.}}:
\batitle{Sex differences in urban commuting patterns}.
\bjtitle{The American economic review}
\bvolume{76}(\bissue{2}),
\bfpage{368}--\blpage{372}
(\byear{1986})
\end{barticle}
\endbibitem

\bibitem[\protect\citeauthoryear{Hanson and Pratt}{1988}]{hanson1988spatial}
\begin{barticle}
\bauthor{\bsnm{Hanson}, \binits{S.}},
\bauthor{\bsnm{Pratt}, \binits{G.}}:
\batitle{Spatial dimensions of the gender division of labor in a local labor market}.
\bjtitle{Urban geography}
\bvolume{9}(\bissue{2}),
\bfpage{180}--\blpage{202}
(\byear{1988})
\end{barticle}
\endbibitem

\bibitem[\protect\citeauthoryear{Macedo et~al.}{2022}]{macedo_differences_2022}
\begin{barticle}
\bauthor{\bsnm{Macedo}, \binits{M.}},
\bauthor{\bsnm{Lotero}, \binits{L.}},
\bauthor{\bsnm{Cardillo}, \binits{A.}},
\bauthor{\bsnm{Menezes}, \binits{R.}},
\bauthor{\bsnm{Barbosa}, \binits{H.}}:
\batitle{Differences in the spatial landscape of urban mobility: {Gender} and socioeconomic perspectives}.
\bjtitle{PLOS ONE}
\bvolume{17}(\bissue{3}),
\bfpage{0260874}
(\byear{2022})
\doiurl{10.1371/journal.pone.0260874}
\end{barticle}
\endbibitem

\bibitem[\protect\citeauthoryear{Collins et~al.}{2024}]{collins2024spatiotemporal}
\begin{barticle}
\bauthor{\bsnm{Collins}, \binits{T.}},
\bauthor{\bsnm{Di~Clemente}, \binits{R.}},
\bauthor{\bsnm{Guti{\'e}rrez-Roig}, \binits{M.}},
\bauthor{\bsnm{Botta}, \binits{F.}}:
\batitle{Spatiotemporal gender differences in urban vibrancy}.
\bjtitle{Environment and Planning B: Urban Analytics and City Science}
\bvolume{51}(\bissue{7}),
\bfpage{1430}--\blpage{1446}
(\byear{2024})
\end{barticle}
\endbibitem

\bibitem[\protect\citeauthoryear{Yasenov}{2020}]{yasenov_who_2020}
\begin{barticle}
\bauthor{\bsnm{Yasenov}, \binits{V.}}:
\batitle{Who {Can} {Work} from {Home}?}
\bjtitle{SSRN Electronic Journal}
(\byear{2020})
\doiurl{10.2139/ssrn.3590895}
\end{barticle}
\endbibitem

\bibitem[\protect\citeauthoryear{Caros and Zhao}{2024}]{caros_need_2024}
\begin{barticle}
\bauthor{\bsnm{Caros}, \binits{N.S.}},
\bauthor{\bsnm{Zhao}, \binits{J.}}:
\batitle{The need for an interdisciplinary approach to remote work and urban policy}.
\bjtitle{Nature Cities}
\bvolume{1}(\bissue{9}),
\bfpage{547}--\blpage{554}
(\byear{2024})
\doiurl{10.1038/s44284-024-00103-y}
\end{barticle}
\endbibitem

\bibitem[\protect\citeauthoryear{Delventhal et~al.}{2022}]{delventhal_jue_2022}
\begin{barticle}
\bauthor{\bsnm{Delventhal}, \binits{M.J.}},
\bauthor{\bsnm{Kwon}, \binits{E.}},
\bauthor{\bsnm{Parkhomenko}, \binits{A.}}:
\batitle{{JUE} {Insight}: {How} do cities change when we work from home?}
\bjtitle{Journal of Urban Economics}
\bvolume{127},
\bfpage{103331}
(\byear{2022})
\doiurl{10.1016/j.jue.2021.103331}
\end{barticle}
\endbibitem

\bibitem[\protect\citeauthoryear{Monte et~al.}{2023}]{monte_remote_2023}
\begin{botherref}
\oauthor{\bsnm{Monte}, \binits{F.}},
\oauthor{\bsnm{Porcher}, \binits{C.}},
\oauthor{\bsnm{Rossi-Hansberg}, \binits{E.}}:
Remote {Work} and {City} {Structure}.
Technical Report w31494,
National Bureau of Economic Research,
Cambridge, MA
(July 2023).
\doiurl{10.3386/w31494}
\end{botherref}
\endbibitem

\bibitem[\protect\citeauthoryear{Gibbs et~al.}{2023}]{gibbs_harnessing_2023}
\begin{botherref}
\oauthor{\bsnm{Gibbs}, \binits{H.}},
\oauthor{\bsnm{Ballantyne}, \binits{P.}},
\oauthor{\bsnm{Cheshire}, \binits{J.}},
\oauthor{\bsnm{Singleton}, \binits{A.}}:
Harnessing mobility data to capture changing work from home behaviours between censuses.
The Geographical Journal
\textbf{190}(2)
(2023)
\doiurl{10.1111/geoj.12555}
\end{botherref}
\endbibitem

\bibitem[\protect\citeauthoryear{Hansen et~al.}{2023}]{hansen_remote_2023}
\begin{botherref}
\oauthor{\bsnm{Hansen}, \binits{S.}},
\oauthor{\bsnm{Lambert}, \binits{P.J.}},
\oauthor{\bsnm{Bloom}, \binits{N.}},
\oauthor{\bsnm{Davis}, \binits{S.}},
\oauthor{\bsnm{Sadun}, \binits{R.}},
\oauthor{\bsnm{Taska}, \binits{B.}}:
Remote {Work} across {Jobs}, {Companies}, and {Space}.
Technical Report w31007,
National Bureau of Economic Research,
Cambridge, MA
(March 2023).
\doiurl{10.3386/w31007} .
\url{http://www.nber.org/papers/w31007.pdf}
\end{botherref}
\endbibitem

\bibitem[\protect\citeauthoryear{T{\'o}th et~al.}{2021}]{toth2021inequality}
\begin{barticle}
\bauthor{\bsnm{T{\'o}th}, \binits{G.}},
\bauthor{\bsnm{Wachs}, \binits{J.}},
\bauthor{\bsnm{Di~Clemente}, \binits{R.}},
\bauthor{\bsnm{Jakobi}, \binits{{\'A}.}},
\bauthor{\bsnm{S{\'a}gv{\'a}ri}, \binits{B.}},
\bauthor{\bsnm{Kert{\'e}sz}, \binits{J.}},
\bauthor{\bsnm{Lengyel}, \binits{B.}}:
\batitle{Inequality is rising where social network segregation interacts with urban topology}.
\bjtitle{Nature communications}
\bvolume{12}(\bissue{1}),
\bfpage{1143}
(\byear{2021})
\end{barticle}
\endbibitem

\bibitem[\protect\citeauthoryear{Aksoy et~al.}{2026}]{aksoy_younger_2026}
\begin{botherref}
\oauthor{\bsnm{Aksoy}, \binits{C.G.}},
\oauthor{\bsnm{Barrero}, \binits{J.M.}},
\oauthor{\bsnm{Bloom}, \binits{N.}},
\oauthor{\bsnm{Cranney}, \binits{K.}},
\oauthor{\bsnm{Davis}, \binits{S.}},
\oauthor{\bsnm{Dolls}, \binits{M.}},
\oauthor{\bsnm{Zarate}, \binits{P.}}:
Younger {Firms} and {CEOs} {Allow} {More} {Work} from {Home}.
Technical Report w34795,
National Bureau of Economic Research,
Cambridge, MA
(February 2026).
\doiurl{10.3386/w34795}
\end{botherref}
\endbibitem

\bibitem[\protect\citeauthoryear{Cullen et~al.}{2025}]{cullen_home_2025}
\begin{barticle}
\bauthor{\bsnm{Cullen}, \binits{Z.}},
\bauthor{\bsnm{Pakzad-Hurson}, \binits{B.}},
\bauthor{\bsnm{Perez-Truglia}, \binits{R.}}:
\batitle{Home {Sweet} {Home}: {How} {Much} {Do} {Employees} {Value} {Remote} {Work}?}
\bjtitle{AEA Papers and Proceedings}
\bvolume{115},
\bfpage{276}--\blpage{281}
(\byear{2025})
\doiurl{10.1257/pandp.20251029}
\end{barticle}
\endbibitem

\bibitem[\protect\citeauthoryear{Yang et~al.}{2021}]{yang_effects_2021}
\begin{barticle}
\bauthor{\bsnm{Yang}, \binits{L.}},
\bauthor{\bsnm{Holtz}, \binits{D.}},
\bauthor{\bsnm{Jaffe}, \binits{S.}},
\bauthor{\bsnm{Suri}, \binits{S.}},
\bauthor{\bsnm{Sinha}, \binits{S.}},
\bauthor{\bsnm{Weston}, \binits{J.}},
\bauthor{\bsnm{Joyce}, \binits{C.}},
\bauthor{\bsnm{Shah}, \binits{N.}},
\bauthor{\bsnm{Sherman}, \binits{K.}},
\bauthor{\bsnm{Hecht}, \binits{B.}},
\bauthor{\bsnm{Teevan}, \binits{J.}}:
\batitle{The effects of remote work on collaboration among information workers}.
\bjtitle{Nature Human Behaviour}
\bvolume{6}(\bissue{1}),
\bfpage{43}--\blpage{54}
(\byear{2021})
\doiurl{10.1038/s41562-021-01196-4}
\end{barticle}
\endbibitem

\bibitem[\protect\citeauthoryear{Brucks and Levav}{2022}]{brucks_virtual_2022}
\begin{barticle}
\bauthor{\bsnm{Brucks}, \binits{M.S.}},
\bauthor{\bsnm{Levav}, \binits{J.}}:
\batitle{Virtual communication curbs creative idea generation}.
\bjtitle{Nature}
\bvolume{605}(\bissue{7908}),
\bfpage{108}--\blpage{112}
(\byear{2022})
\doiurl{10.1038/s41586-022-04643-y}
\end{barticle}
\endbibitem

\bibitem[\protect\citeauthoryear{Bernstein et~al.}{2018}]{bernstein_how_2018}
\begin{barticle}
\bauthor{\bsnm{Bernstein}, \binits{E.}},
\bauthor{\bsnm{Shore}, \binits{J.}},
\bauthor{\bsnm{Lazer}, \binits{D.}}:
\batitle{How intermittent breaks in interaction improve collective intelligence}.
\bjtitle{Proceedings of the National Academy of Sciences}
\bvolume{115}(\bissue{35}),
\bfpage{8734}--\blpage{8739}
(\byear{2018})
\doiurl{10.1073/pnas.1802407115}
\end{barticle}
\endbibitem

\bibitem[\protect\citeauthoryear{Bloom et~al.}{2024}]{bloom_hybrid_2024}
\begin{barticle}
\bauthor{\bsnm{Bloom}, \binits{N.}},
\bauthor{\bsnm{Han}, \binits{R.}},
\bauthor{\bsnm{Liang}, \binits{J.}}:
\batitle{Hybrid working from home improves retention without damaging performance}.
\bjtitle{Nature}
\bvolume{630}(\bissue{8018}),
\bfpage{920}--\blpage{925}
(\byear{2024})
\doiurl{10.1038/s41586-024-07500-2}
\end{barticle}
\endbibitem

\bibitem[\protect\citeauthoryear{Baker et~al.}{2026}]{baker_shopping_2026}
\begin{botherref}
\oauthor{\bsnm{Baker}, \binits{S.}},
\oauthor{\bsnm{Bloom}, \binits{N.}},
\oauthor{\bsnm{Johnson}, \binits{S.}},
\oauthor{\bsnm{Obradović}, \binits{J.}}:
Shopping {From} {Home}.
Technical Report w34883,
National Bureau of Economic Research,
Cambridge, MA
(February 2026).
\doiurl{10.3386/w34883}
\end{botherref}
\endbibitem

\bibitem[\protect\citeauthoryear{Rossi~Mori et~al.}{2025}]{rossi2025time}
\begin{barticle}
\bauthor{\bsnm{Rossi~Mori}, \binits{L.}},
\bauthor{\bsnm{Loreto}, \binits{V.}},
\bauthor{\bsnm{Di~Clemente}, \binits{R.}}:
\batitle{Time-space dynamics of income segregation in the city of milan}.
\bjtitle{Pnas Nexus}
\bvolume{4}(\bissue{9}),
\bfpage{283}
(\byear{2025})
\end{barticle}
\endbibitem

\end{thebibliography}

\clearpage
\begin{appendices}
\renewcommand{\thefigure}{S\arabic{figure}}
\renewcommand{\thetable}{S\arabic{table}}
\renewcommand{\thesection}{S\arabic{section}}

\setcounter{figure}{0}
\setcounter{table}{0}
\setcounter{section}{0}

\section{Spatial and temporal smoothing}
\label{si:sec:smoothing}

\subsection{Spatial smoothing}
\label{si:sec:spatial-smoothing}
Raw mobile phone location counts (workplace presence, home presence, and general traffic)
were spatially smoothed at H3 resolution-10 to reduce measurement noise arising from
individual-level geolocation error. For each hexagon, we computed a spatial moving average
over its immediate ring-1 neighbourhood, summing counts across the neighbourhood and
dividing by seven; the number of neighbours.

Two observation periods were used as a quasi-natural experiment: an unrestricted
\emph{opening} period (1 September -- 31 October 2020) and a strict \emph{curfew} period
(11 November -- 23 December 2020). Two calendar days were identified as anomalous and
excluded from their respective periods prior to any further processing: 23 October (opening)
and 12 December (curfew, a Saturday that was a workday). 

\subsection{Temporal smoothing}
\label{si:sec:temporal-smoothing}

Each period's spatially-smoothed hourly data were collapsed into a representative 48-hour
sequence per hexagon: hours 0--23 for weekdays and hours 24--47 for weekends, each hour
being the mean across all matching weekday or weekend observations within the period.

A location-level regularity criterion was then applied to distinguish real workplaces
and homes. Therefore, differentiating between locations with no work activity, to those with numbers too low to include due to privacy protection. A hexagon's workplace (or home) values were retained only if that hexagon showed non-zero presence in \emph{every} week of the observation period; hexagons that failed this criterion had their entire workplace or home series set to missing, rather than being interpreted as a location with no activity at that hour.

For hexagons passing this criterion, a Gaussian smoothing kernel ($\sigma = 1$) was applied across each hexagon's own 48-hour sequence to reduce hour-to-hour sampling noise, after adding a pseudocount of 1 to avoid zero values ahead of the subsequent log transform. The resulting Gaussian-smoothed values were log-transformed, and then z-scored across each hexagon's own 48-hour sequence, giving a measure of how each hour of the day compares to that specific location's average activity level.

\clearpage
\section{Descriptive statistics and correlations}
\label{si:sec:descriptive}

\subsection{Summary statistics}
\label{si:sec:summary-stats}

Table S1 reports descriptive statistics (N, mean, standard deviation, minimum, 25th/50th/75th percentiles, and maximum) for the main variables used in the regression analysis and in Figure~2 (the cluster radar plot), computed on the absolute values for interpretability. Table S2 reports the same statistics on the standardized scale used directly in the regression models. 

\begin{table}[b]
\centering
\caption{Summary statistics of main variables - Absolute values}
\label{tab:summary_stats_absolute}
\begin{tabular}{lrrrrrrrr}
\toprule
 & N & Mean & SD & Min & P25 & Median & P75 & Max \\
Variable &  &  &  &  &  &  &  &  \\
\midrule
WPD & 8861 & -0.680 & 3.950 & -36.930 & -1.470 & 0.110 & 1.080 & 13.660 \\
Work activity & 8861 & 15.520 & 18.750 & 1.020 & 2.770 & 8.030 & 20.420 & 156.460 \\
Productivity & 8861 & 30360.630 & 58963.050 & 0.000 & 9284.930 & 17741.440 & 33673.290 & 1826233.830 \\
Nr. of companies & 8861 & 3.310 & 5.910 & 1.000 & 1.000 & 2.000 & 3.000 & 143.000 \\
Industry entropy & 8861 & 0.920 & 1.110 & 0.000 & 0.000 & 0.920 & 1.580 & 5.950 \\
Distance to centre & 8861 & 7565.890 & 3952.620 & 56.350 & 4502.240 & 7462.980 & 10080.040 & 20648.820 \\
Income entropy & 8861 & 0.980 & 0.220 & 0.000 & 0.900 & 0.990 & 1.080 & 1.580 \\
Ratio of women & 8861 & 0.300 & 0.040 & 0.100 & 0.270 & 0.300 & 0.320 & 0.490 \\
Ratio of high income & 8861 & 0.020 & 0.040 & 0.000 & 0.000 & 0.000 & 0.020 & 1.000 \\
\bottomrule
\end{tabular}
\end{table}

\begin{table}
\centering
\caption{Summary statistics of main variables - Standardized values}
\begin{tabular}{lrrrrrrrr}
\toprule
 & N & Mean & SD & Min & P25 & Median & P75 & Max \\
Variable &  &  &  &  &  &  &  &  \\
\midrule
Workplace dependence & 8861 & 0.000 & 1.000 & -11.690 & 0.060 & 0.130 & 0.260 & 4.510 \\
Work activity & 8861 & -0.000 & 1.000 & -0.600 & -0.600 & -0.450 & 0.130 & 8.730 \\
Productivity & 8861 & 0.000 & 1.000 & -0.510 & -0.360 & -0.210 & 0.060 & 30.460 \\
Number of companies & 8861 & 0.000 & 1.000 & -0.390 & -0.390 & -0.220 & -0.050 & 23.650 \\
Industry entropy & 8861 & -0.000 & 1.000 & -0.830 & -0.830 & -0.000 & 0.600 & 4.540 \\
Distance to centre & 8861 & -0.000 & 1.000 & -1.900 & -0.780 & -0.030 & 0.640 & 3.310 \\
Income entropy & 8861 & -0.000 & 1.000 & -3.930 & -0.320 & 0.060 & 0.440 & 2.460 \\
Ratio of women & 8861 & 0.000 & 1.000 & -3.690 & -0.260 & 0.150 & 0.540 & 2.730 \\
Ratio of high income & 8861 & -0.000 & 1.000 & -0.480 & -0.480 & -0.480 & 0.090 & 27.560 \\
\bottomrule
\end{tabular}
\end{table}

\clearpage
\subsection{Distributions}

\begin{figure}[H]
\centering
\includegraphics[width=\linewidth]{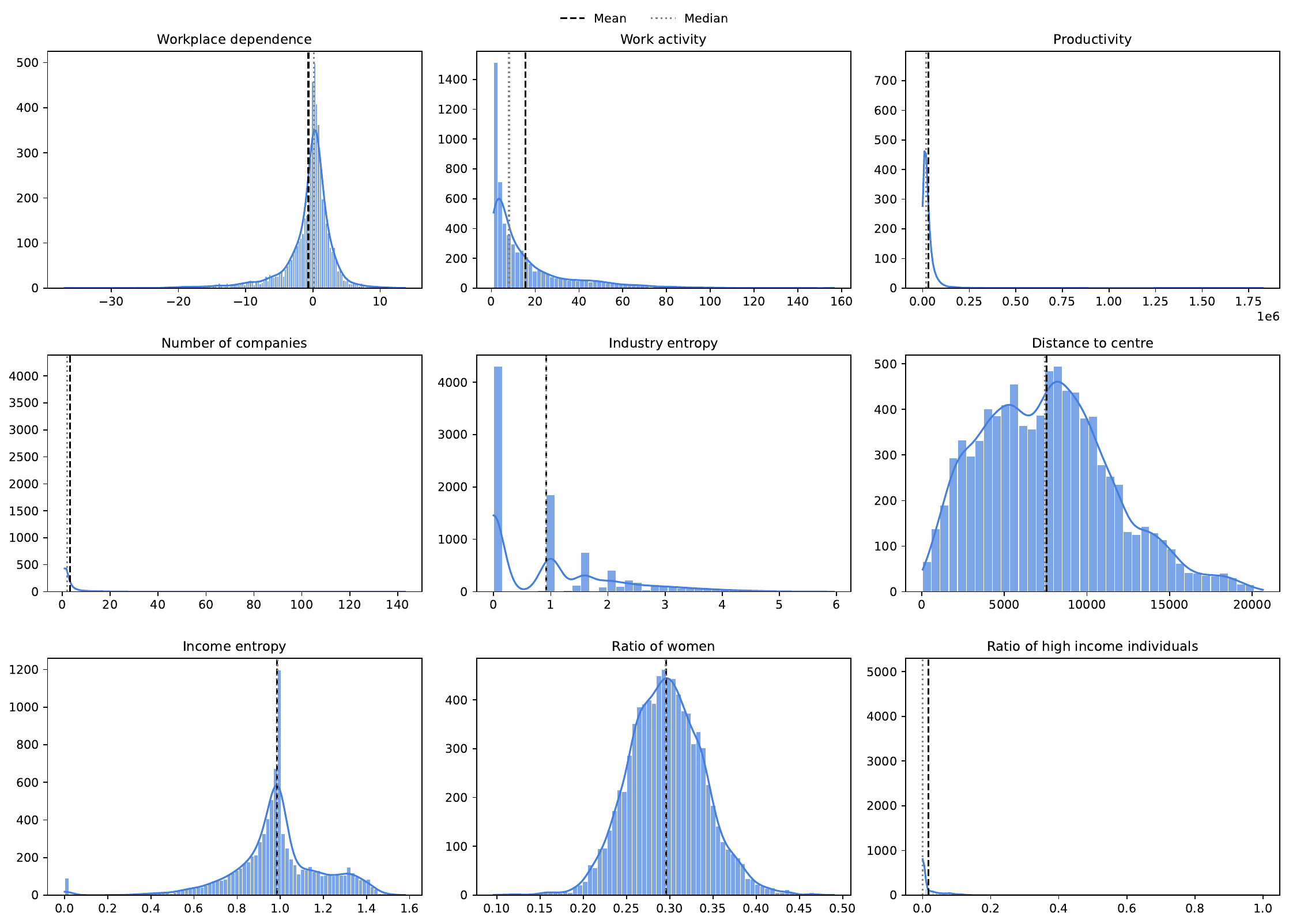}
\caption{Distributions of the main variables used in the regression analysis.}
\label{si:fig:variable-distributions}
\end{figure}

Figure~\ref{si:fig:variable-distributions} shows the marginal distribution (histogram with
kernel density estimate) of each of the main variables listed above, on the raw scale.

\begin{figure}[H]
\centering
\includegraphics[width=0.8\linewidth]{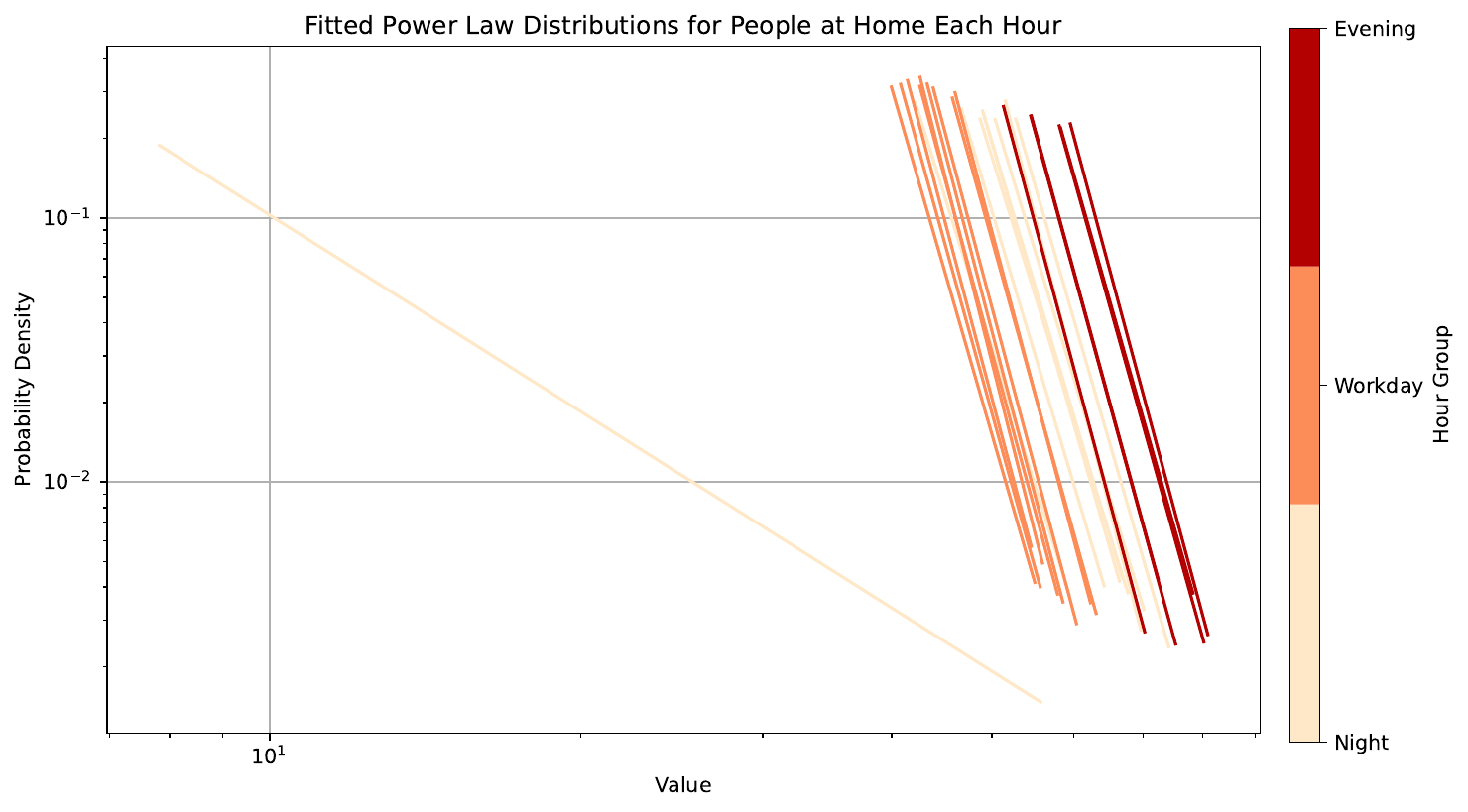}
\caption{Fitted power-law distributions of home presence, by weekday hour.}
\label{si:fig:powerlaw-home}
\end{figure}

\begin{figure}[H]
\centering
\includegraphics[width=0.8\linewidth]{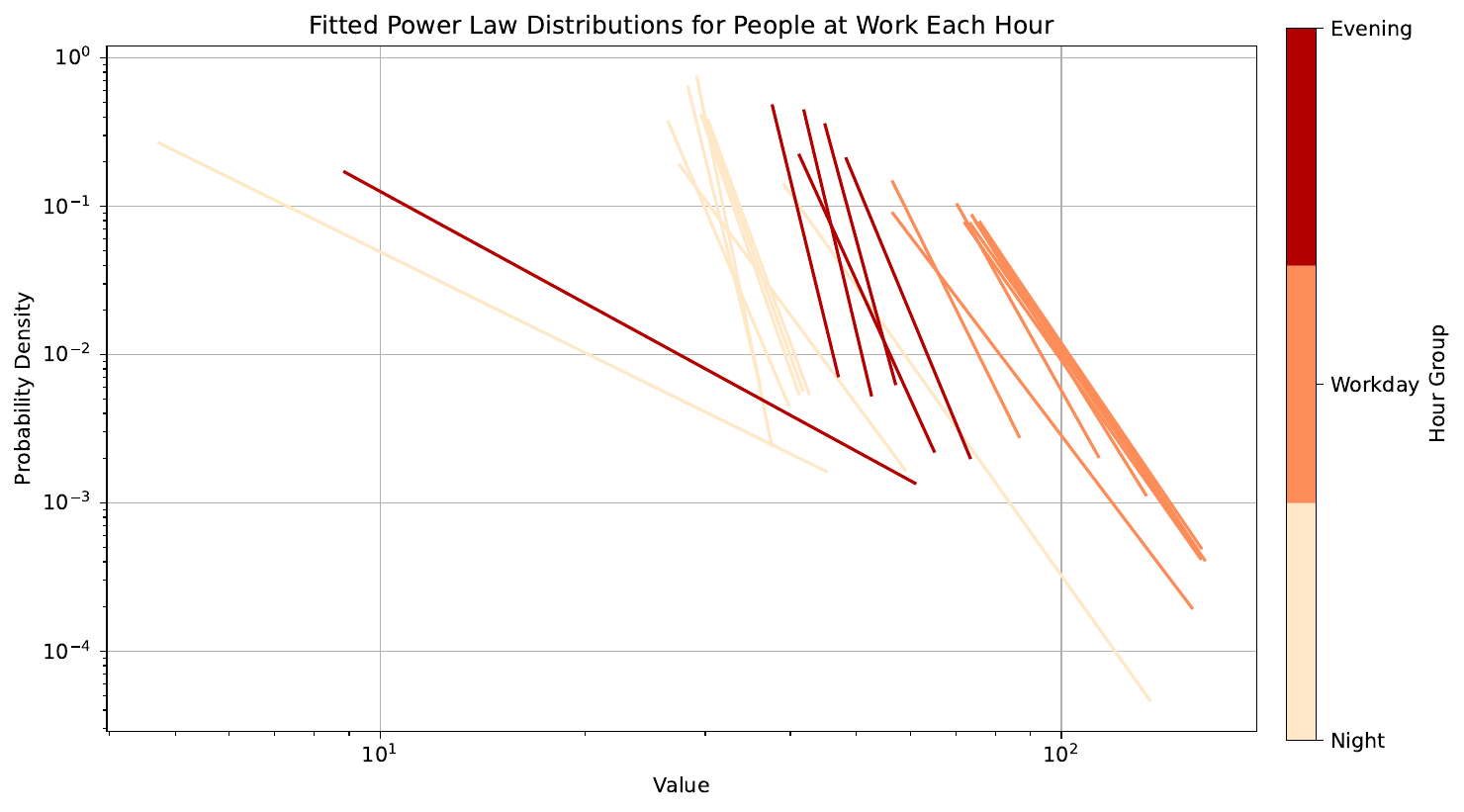}
\caption{Fitted power-law distributions of workplace presence, by weekday hour.}
\label{si:fig:powerlaw-work}
\end{figure}

\begin{figure}[H]
\centering
\includegraphics[width=0.8\linewidth]{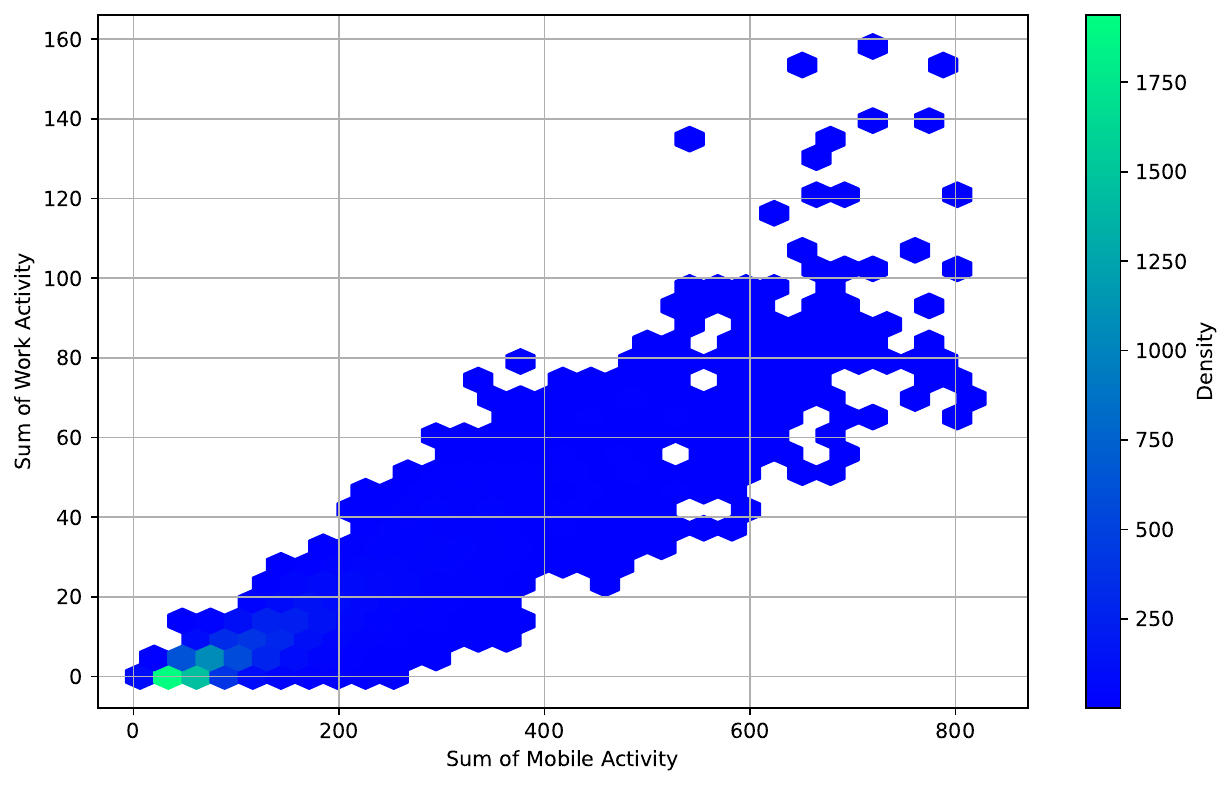}
\caption{Hexbin density plot of traffic versus workplace activity per hexagon (weekday hours, opening period).}
\label{si:fig:hexbin}
\end{figure}

\begin{figure}[H]
\centering
\includegraphics[width=0.8\linewidth]{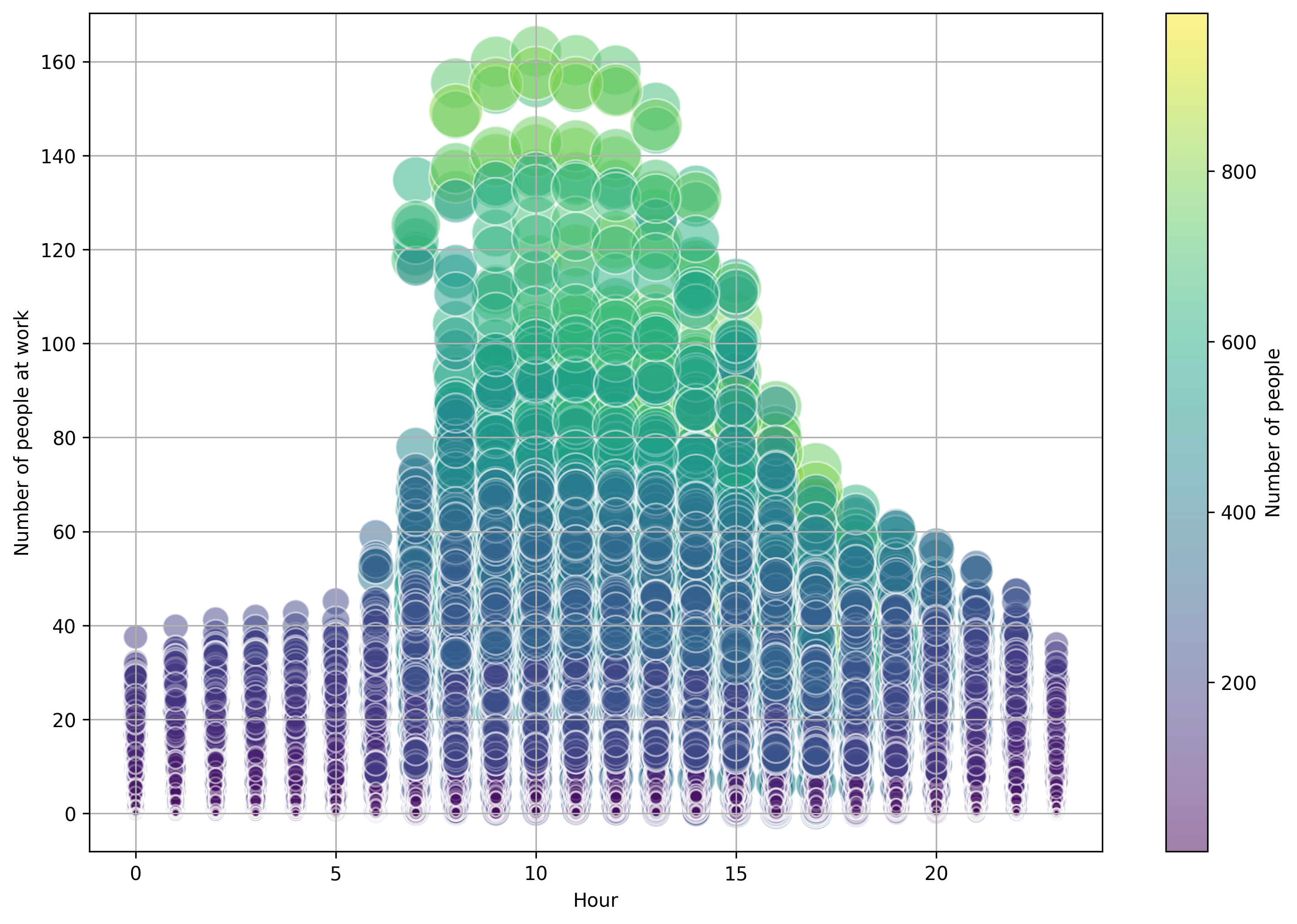}
\caption{Workplace activity across the 48-hour cycle, bubble size and colour scaled to traffic volume.}
\label{si:fig:bubble}
\end{figure}

Figures~\ref{si:fig:powerlaw-home} and \ref{si:fig:powerlaw-work} show fitted power-law
distributions of home and workplace presence respectively, estimated separately for each
weekday hour (0--23), coloured by broad time-of-day grouping (night, workday, evening).
Figure~\ref{si:fig:hexbin} shows the joint distribution of workplace activity and general
traffic per hexagon (weekday hours only) as a hexbin density plot, and
Figure~\ref{si:fig:bubble} shows workplace activity across the 48-hour cycle as a bubble
plot, with bubble size and colour both scaled to traffic volume.

\subsection{Correlations}
\label{si:sec:correlations}

\begin{figure}[H]
\centering
\includegraphics[width=0.8\linewidth]{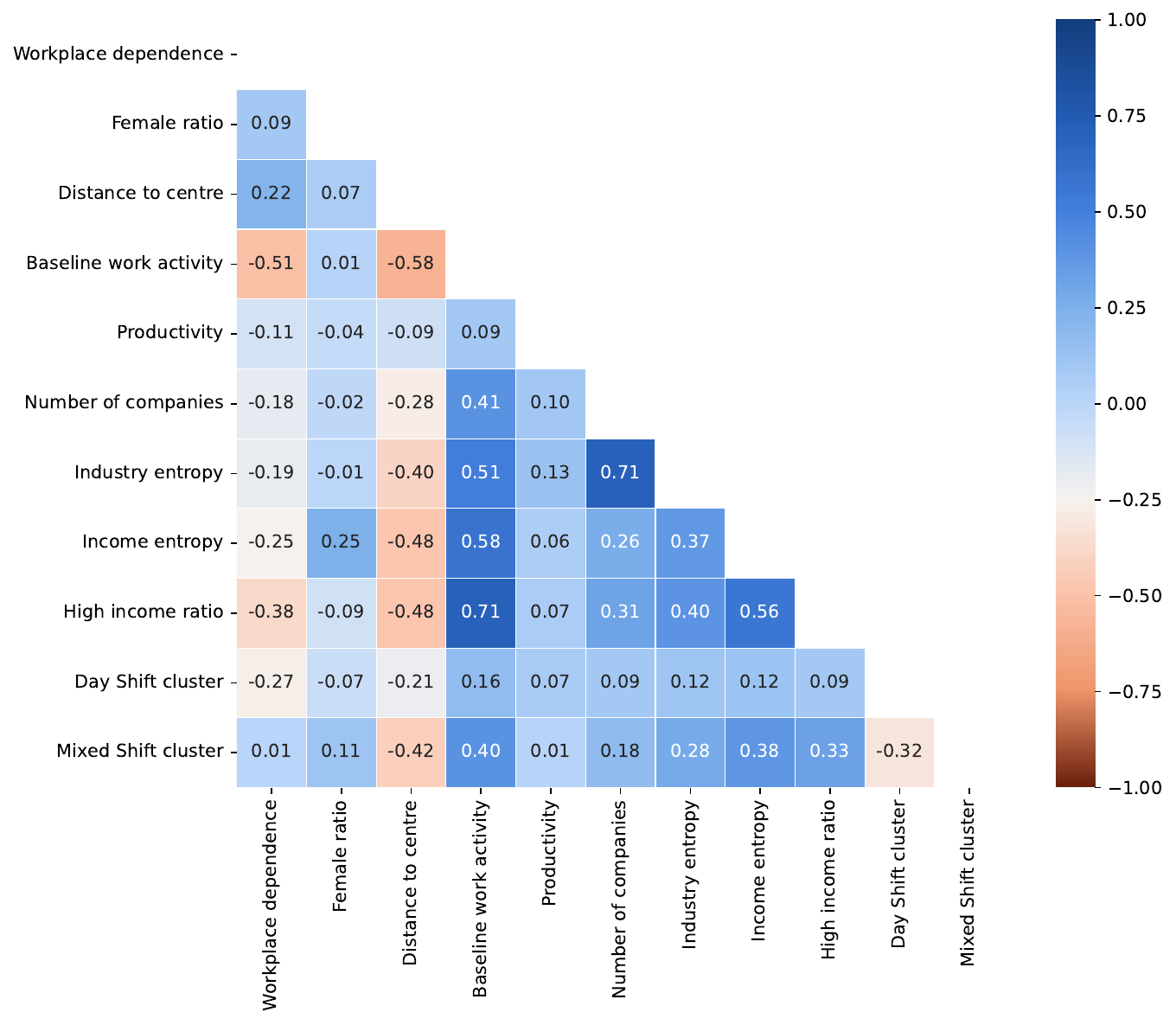}
\caption{Correlation matrix of the main regression variables.}
\label{si:fig:corr-matrix}
\end{figure}

Figure~\ref{si:fig:corr-matrix} shows pairwise Pearson correlations between the main
variables used in the regression analysis.

\subsection{Spatial distributions}
\label{si:sec:maps}

\begin{figure}[H]
\centering
\includegraphics[width=0.7\linewidth]{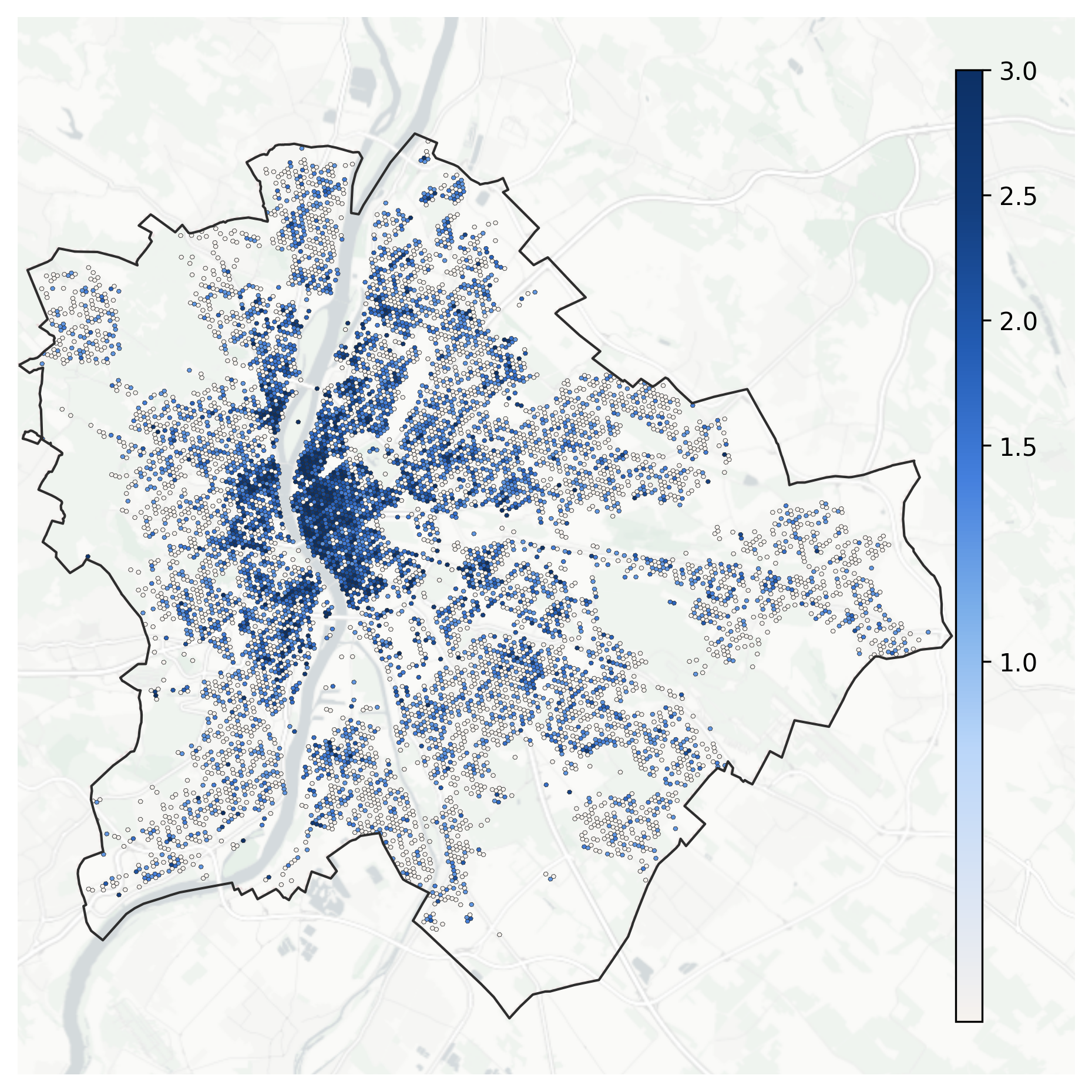}
\caption{Log-scaled distribution of the number of registered companies per hexagon.}
\label{si:fig:map-companies}
\end{figure}

\begin{figure}[H]
\centering
\includegraphics[width=0.7\linewidth]{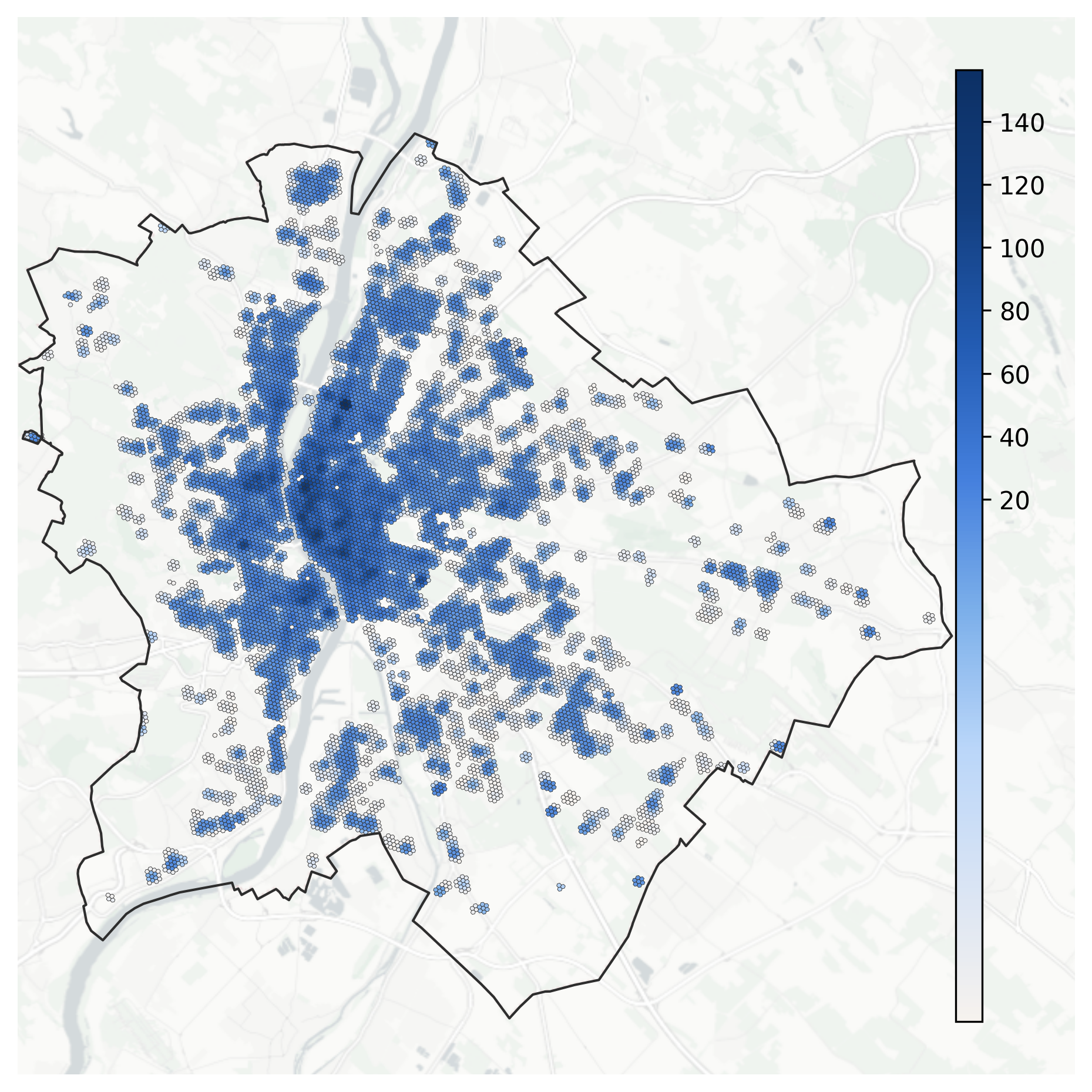}
\caption{Workplace activity per hexagon, opening period.}
\label{si:fig:map-activity}
\end{figure}

\begin{figure}[H]
\centering
\includegraphics[width=0.7\linewidth]{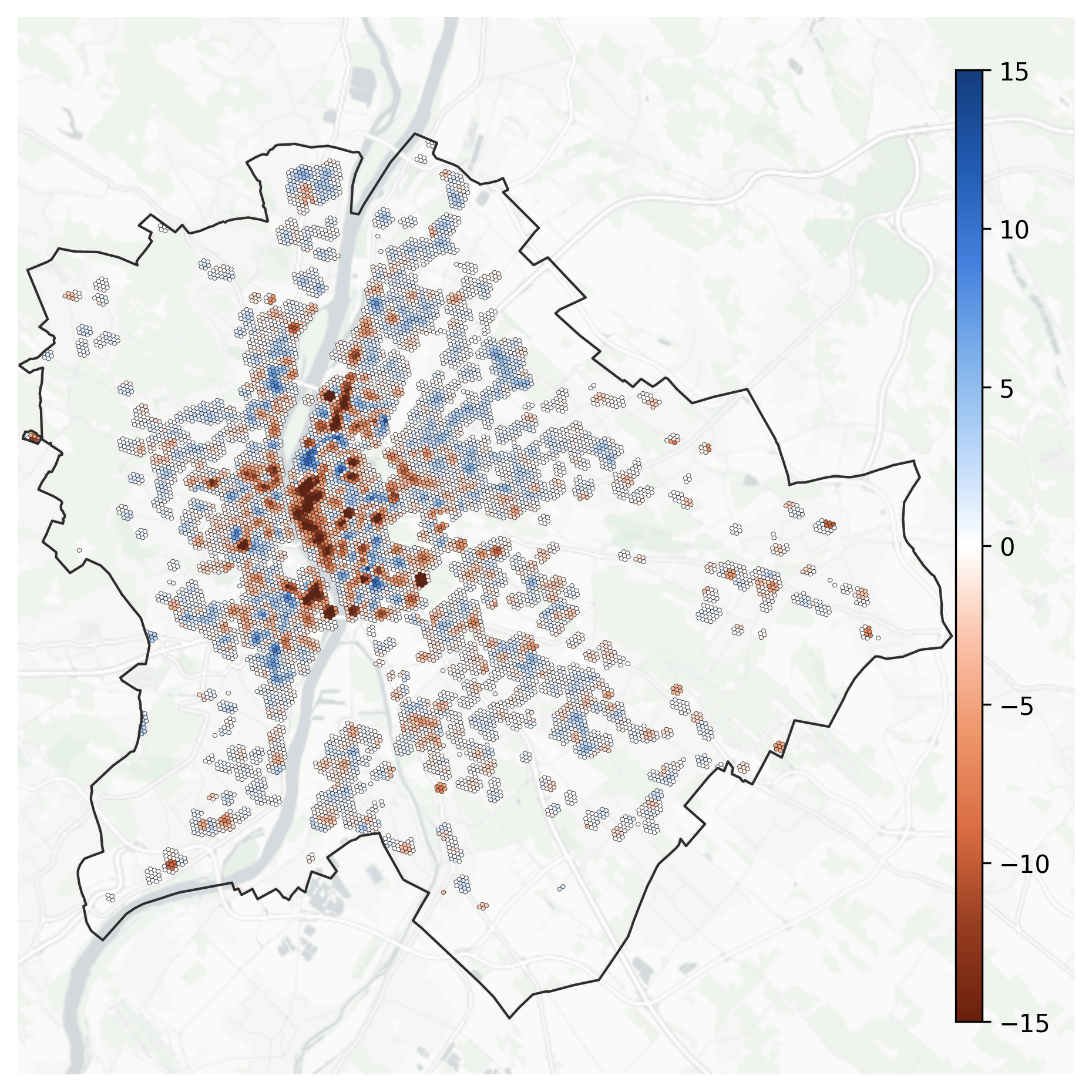}
\caption{Workplace dependence per hexagon.}
\label{si:fig:map-wpd}
\end{figure}

Figures~\ref{si:fig:map-companies}--\ref{si:fig:map-wpd} show the spatial distribution
across Budapest of, respectively, the number of registered companies per hexagon (on a log
scale), average workplace activity during the opening period, and workplace dependence. 

\clearpage
\section{Clustering}
\label{si:sec:clustering}

\subsection{Elbow method}
\label{si:sec:elbow}

\begin{figure}[h]
\centering
\includegraphics[width=0.7\linewidth]{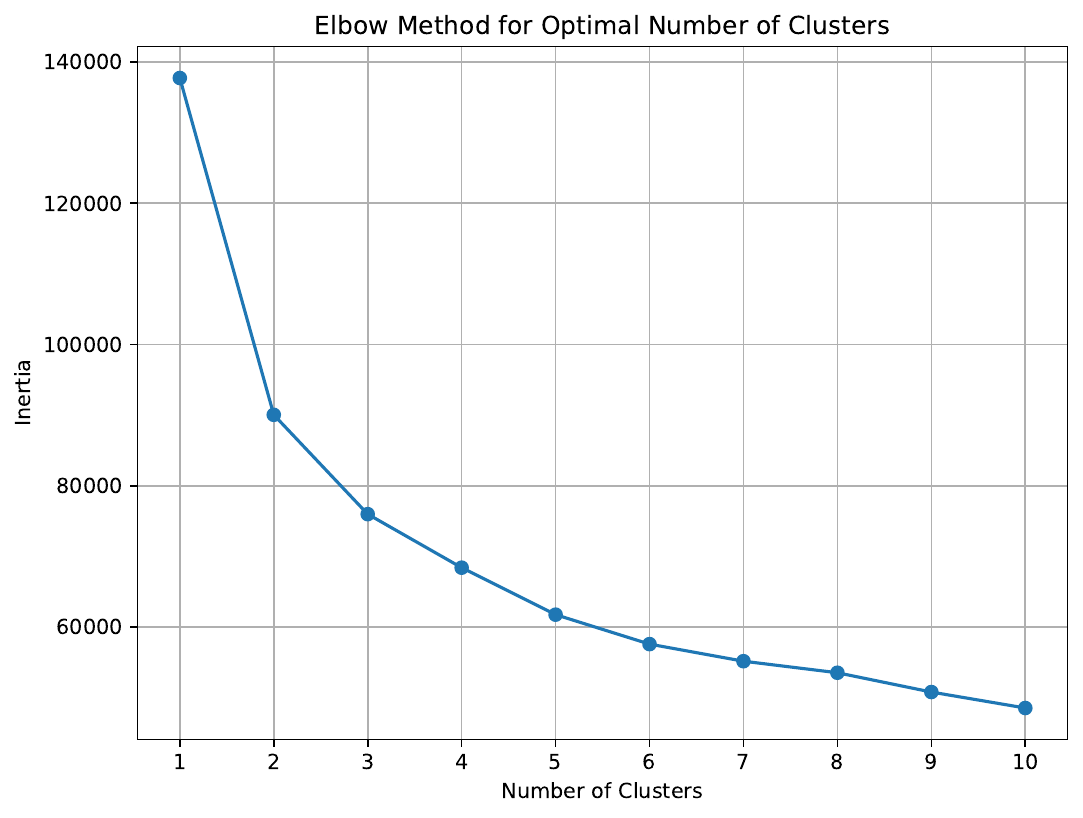}
\caption{Elbow plot for k-means clustering of workplace activity temporal profiles.}
\label{si:fig:elbow}
\end{figure}

Figure~\ref{si:fig:elbow} shows the k-means elbow plot (inertia against number of clusters, $k=1$ to $10$) used to select the number of temporal work-activity clusters, computed on the z-scored, log-transformed workplace activity sequence.

\subsection{Temporal cluster with mean}
\label{si:sec:cluster-profiles}

\begin{figure}[h]
\centering
\includegraphics[width=\linewidth]{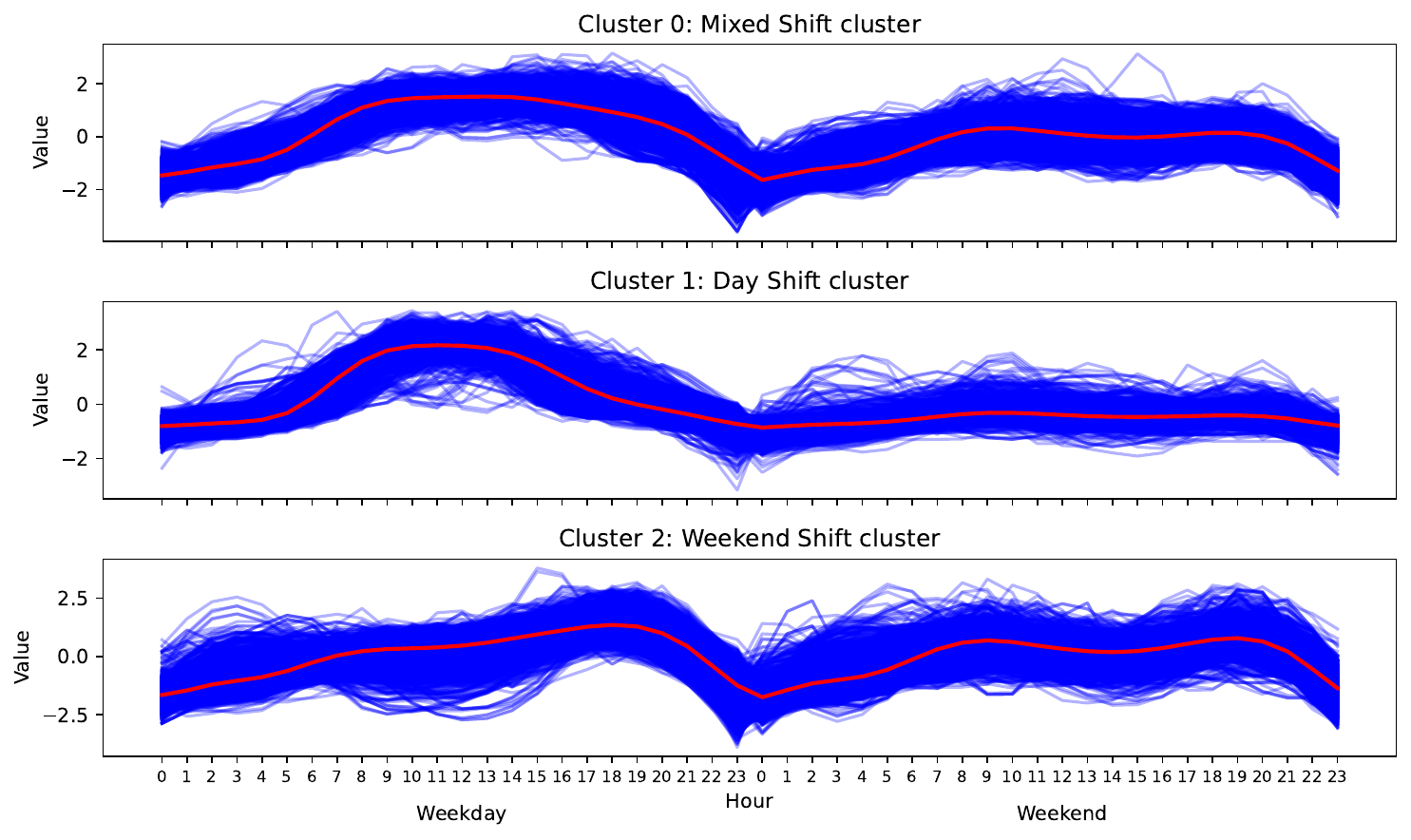}
\caption{Temporal profile of workplace activity by cluster (cluster means with individual hexagon sequences).}
\label{si:fig:cluster-temporal}
\end{figure}

Figure~\ref{si:fig:cluster-temporal} shows the mean 48-hour workplace activity profile for each of the three identified clusters, with individual hexagon sequences with cluster mean.

\clearpage
\section{TE\'AOR industry sector classification}
\label{si:sec:teaor}

Table~\ref{si:tab:teaor} lists the correspondence between the TE\'AOR'08 (Hungarian
activity classification, equivalent to NACE Rev.\,2) one-letter sections used as industry
fixed effects throughout the regression analysis, their full sector names, and the
corresponding two-digit TE\'AOR'08 divisions each section aggregates. Sections T (activities
of households as employers) and U (activities of extraterritorial organizations) are part
of the full TE\'AOR'08 classification but do not appear in the analysis sample and are
omitted here.

\begin{table}[h]
\centering
\begin{tabularx}{\textwidth}{cXc}
\toprule
Sector & Name & TE\'AOR divisions \\
\midrule
A & Agriculture, forestry and fishing & 01--03 \\
B & Mining and quarrying & 05--09 \\
C & Manufacturing & 10--33 \\
D & Electricity, gas, steam and air conditioning supply & 35 \\
E & Water supply; sewerage, waste management and remediation activities & 36--39 \\
F & Construction & 41--43 \\
G & Wholesale and retail trade; repair of motor vehicles and motorcycles & 45--47 \\
H & Transportation and storage & 49--53 \\
I & Accommodation and food service activities & 55--56 \\
J & Information and communication & 58--63 \\
K & Financial and insurance activities & 64--66 \\
L & Real estate activities & 68 \\
M & Professional, scientific and technical activities & 69--75 \\
N & Administrative and support service activities & 77--82 \\
O & Public administration and defence; compulsory social security & 84 \\
P & Education & 85 \\
Q & Human health and social work activities & 86--88 \\
R & Arts, entertainment and recreation & 90--93 \\
S & Other service activities & 94--96 \\
\bottomrule
\end{tabularx}
\caption{TE\'AOR'08 sector classification used as industry fixed effects.}
\label{si:tab:teaor}
\end{table}

\clearpage
\section{Regression tables}
\label{si:sec:regression}

Table S4 reports all four OLS specifications side by side: the main
specification underlying Figure~1 (column 1), the two distance-interaction specifications underlying Figure~4b--c (columns 2--3), and the single-company robustness check (column 4). The robustness check reports the results restricted to hexagons with exactly one registered company. The full industry fixed effects are replaced with indicators for the five sectors best represented among single-company locations (professional/scientific, administrative, arts/entertainment, human health, and financial/insurance), with wholesale/retail retained as the reference category as in the main specification.
Standard errors are heteroskedasticity-robust (HC3) throughout.

\begin{table}[b]
\caption{OLS Regression}
\label{si:tab:ols}
\begin{tabular}{lllll}
\hline
                                         & Baseline   & Female Interact & Income Interact & Single-company  \\
\hline
Intercept                                & -0.0066    & 0.0040          & -0.0461**       & -0.0961***      \\
                                         & (0.0177)   & (0.0176)        & (0.0200)        & (0.0337)        \\
Baseline work activity             & -0.6114*** & -0.6248***      & -0.6650***      & -0.5297***      \\
                                         & (0.0319)   & (0.0314)        & (0.0341)        & (0.0657)        \\
Productivity                             & -0.0591*** & -0.0560***      & -0.0580***      & -0.0162         \\
                                         & (0.0163)   & (0.0159)        & (0.0159)        & (0.0160)        \\
Number of companies                             & -0.0258    & -0.0244         & -0.0253         &                 \\
                                         & (0.0239)   & (0.0235)        & (0.0238)        &                 \\
Industry entropy                                  & 0.1020***  & 0.1015***       & 0.0956***       &                 \\
                                         & (0.0179)   & (0.0177)        & (0.0176)        &                 \\
Distance to centre                       & -0.1094*** & -0.1234***      & -0.1267***      & -0.0669***      \\
                                         & (0.0092)   & (0.0096)        & (0.0111)        & (0.0091)        \\
High income ratio                  & -0.0617    & -0.0559         & -0.0963*        & -0.0962         \\
                                         & (0.0435)   & (0.0418)        & (0.0561)        & (0.0799)        \\
Income entropy                       & 0.0090     & 0.0257*         & 0.0532**        & -0.0141         \\
                                         & (0.0137)   & (0.0135)        & (0.0238)        & (0.0176)        \\
Female ratio                    & 0.0674***  & 0.1346***       & 0.0842***       & 0.0355***       \\
                                         & (0.0073)   & (0.0107)        & (0.0074)        & (0.0090)        \\
DistxFemale ratio &            & -0.1041***      &                 &                 \\
                                         &            & (0.0087)        &                 &                 \\
DistxIncome entropy    &            &                 & -0.0922***      &                 \\
                                         &            &                 & (0.0170)        &                 \\
R-squared                                & 0.3539     & 0.3630          & 0.3595          & 0.2999          \\
R-squared Adj.                           & 0.3519     & 0.3609          & 0.3573          & 0.2977          \\
Industry FE                              & Yes        & Yes             & Yes             & No              \\
N                                        & 8861       & 8861            & 8861            & 4182            \\
\hline
\end{tabular}
Standard errors in parentheses. \newline 
$*$ p$<$.1, $**$ p$<$.05, $***$p$<$.01
\end{table}

\clearpage

\begin{table}
\caption{Stepwise regression}
\label{si:tab:step}
\begin{tabular}{llllll}
\hline
                                     & Baseline & + City  & + Company & + Socioecon. & + Industry FE  \\
\hline
Intercept                            & 0.0000      & -0.0619*** & -0.0545***   & -0.0441***      & -0.0066           \\
                                     & (0.0091)    & (0.0127)   & (0.0126)     & (0.0122)        & (0.0177)          \\
Baseline work activity         & -0.5150***  & -0.6167*** & -0.6433***   & -0.6095***      & -0.6114***        \\
                                     & (0.0220)    & (0.0253)   & (0.0263)     & (0.0319)        & (0.0319)          \\
Day Shift cluster             &             & -0.3657*** & -0.3705***   & -0.3807***      & -0.3806***        \\
                                     &             & (0.0262)   & (0.0263)     & (0.0273)        & (0.0273)          \\
Mixed Shift cluster             &             & 0.3575***  & 0.3381***    & 0.3125***       & 0.3106***         \\
                                     &             & (0.0227)   & (0.0222)     & (0.0214)        & (0.0215)          \\
Distance to centre                   &             & -0.0959*** & -0.0878***   & -0.1038***      & -0.1094***        \\
                                     &             & (0.0089)   & (0.0086)     & (0.0090)        & (0.0092)          \\
Productivity                         &             &            & -0.0576***   & -0.0553***      & -0.0591***        \\
                                     &             &            & (0.0162)     & (0.0162)        & (0.0163)          \\
Number of companies                         &             &            & -0.0306      & -0.0270         & -0.0258           \\
                                     &             &            & (0.0241)     & (0.0239)        & (0.0239)          \\
Industry entropy                              &             &            & 0.1026***    & 0.1021***       & 0.1020***         \\
                                     &             &            & (0.0181)     & (0.0179)        & (0.0179)          \\
High income ratio              &             &            &              & -0.0631         & -0.0617           \\
                                     &             &            &              & (0.0434)        & (0.0435)          \\
Income entropy                   &             &            &              & 0.0083          & 0.0090            \\
                                     &             &            &              & (0.0136)        & (0.0137)          \\
Female ratio                &             &            &              & 0.0687***       & 0.0674***         \\
                                     &             &            &              & (0.0073)        & (0.0073)          \\
Agriculture, forestry and fishing &             &            &              &                 & -0.1000           \\
                                     &             &            &              &                 & (0.0834)          \\
Mining and quarrying &             &            &              &                 & 0.2704            \\
                                     &             &            &              &                 & (0.3500)          \\
Manufacturing &             &            &              &                 & -0.0347           \\
                                     &             &            &              &                 & (0.0262)          \\
Electricity, gas, steam and air con &             &            &              &                 & 0.3149*           \\
                                     &             &            &              &                 & (0.1910)          \\
Water supply; sewerage, waste &             &            &              &                 & -0.0575           \\
                                     &             &            &              &                 & (0.1160)          \\
Construction &             &            &              &                 & -0.0267           \\
                                     &             &            &              &                 & (0.0262)          \\
Transportation and storage &             &            &              &                 & -0.0012           \\
                                     &             &            &              &                 & (0.0427)          \\
Accommodation and food service &             &            &              &                 & -0.0360           \\
                                     &             &            &              &                 & (0.0384)          \\
Information and communication &             &            &              &                 & -0.1076**         \\
                                     &             &            &              &                 & (0.0506)          \\
Financial and insurance &             &            &              &                 & -0.2076           \\
                                     &             &            &              &                 & (0.1349)          \\
Real estate activities &             &            &              &                 & 0.0460            \\
                                     &             &            &              &                 & (0.0553)          \\
Professional, scientific and technical &             &            &              &                 & -0.0581*          \\
                                     &             &            &              &                 & (0.0307)          \\
Administrative and support service &             &            &              &                 & -0.1151***        \\
                                     &             &            &              &                 & (0.0419)          \\
Public administration and defence &             &            &              &                 & 2.5465            \\
                                     &             &            &              &                 & (1.7271)          \\
Education &             &            &              &                 & -0.1910           \\
                                     &             &            &              &                 & (0.1676)          \\
Human health and social work &             &            &              &                 & -0.1216***        \\
                                     &             &            &              &                 & (0.0457)          \\
Arts, entertainment and recreation &             &            &              &                 & -0.2097***        \\
                                     &             &            &              &                 & (0.0778)          \\
Other service activities &             &            &              &                 & 0.0649            \\
                                     &             &            &              &                 & (0.0653)          \\
R-squared                            & 0.2652      & 0.3341     & 0.3419       & 0.3495          & 0.3539            \\
R-squared Adj.                       & 0.2651      & 0.3338     & 0.3414       & 0.3487          & 0.3519            \\
N                                    & 8861        & 8861       & 8861         & 8861            & 8861              \\
R2                                   & 0.265       & 0.334      & 0.342        & 0.349           & 0.354             \\
\hline
\end{tabular}
Standard errors in parentheses. \newline 
$*$ p$<$.1, $**$ p$<$.05, $***$p$<$.01
\end{table}

\clearpage
\subsection{SEM regression}
\label{si:sec:sem-regression}

Table S6 reports spatial error model (SEM) estimates corresponding to each of the three OLS specifications in Table S4 (columns 1--3). Spatial weights were constructed from H3 hexagon adjacency: for each hexagon, neighbours were first sought among its immediate ring-1 neighbours; where none of these were present in the estimation sample, the search radius was expanded up to a maximum of ring 5. Hexagons with no in-sample neighbour (4 hexagons) within five rings were dropped from the estimation sample entirely, since a spatial error model requires a fully connected weights matrix.

\begin{table}[b]
\caption{SEM Regression}
\label{si:tab:sem}
\begin{tabular}{lccc}
\toprule
 & Baseline & Female Interact & Income Interact \\
\midrule
Constant & -0.032* & -0.028 & -0.065*** \\
 & (0.018) & (0.018) & (0.019) \\
Baseline work activity & -0.684*** & -0.690*** & -0.724*** \\
 & (0.015) & (0.015) & (0.016) \\
Productivity & -0.008* & -0.008 & -0.008* \\
 & (0.005) & (0.005) & (0.005) \\
Number of companies & -0.010 & -0.010 & -0.010 \\
 & (0.007) & (0.007) & (0.007) \\
Industry entropy & 0.032*** & 0.031*** & 0.030*** \\
 & (0.008) & (0.008) & (0.008) \\
Distance to centre & -0.167*** & -0.177*** & -0.166*** \\
 & (0.016) & (0.016) & (0.016) \\
Day Shift cluster & -0.151*** & -0.148*** & -0.150*** \\
 & (0.022) & (0.022) & (0.022) \\
Mixed Shift cluster & 0.145*** & 0.139*** & 0.139*** \\
 & (0.019) & (0.019) & (0.019) \\
High income ratio & -0.038*** & -0.037*** & -0.053*** \\
 & (0.010) & (0.010) & (0.010) \\
Income entropy & 0.003 & 0.009 & 0.051*** \\
 & (0.010) & (0.010) & (0.012) \\
Female ratio & 0.043*** & 0.089*** & 0.051*** \\
 & (0.008) & (0.011) & (0.008) \\
$\lambda$ & 0.615*** & 0.615*** & 0.615*** \\
 & (0.007) & (0.007) & (0.007) \\
Distance x Female ratio &  & -0.055*** &  \\
 &  & (0.009) &  \\
Distance x Income entropy &  &  & -0.070*** \\
 &  &  & (0.009) \\
\midrule
Industry FE & Yes & Yes & Yes \\
N & 8857 & 8857 & 8857 \\
Pseudo R$^2$ & 0.322 & 0.331 & 0.328 \\
\bottomrule
\end{tabular}
\end{table}

\clearpage
\subsection{Moran's I}
\label{si:sec:moran}

Table~\ref{si:tab:moran} reports Moran's I statistics for the residuals of each OLS
specification, computed under H3 hexagon contiguity weights (see
Section~\ref{si:sec:sem-regression} for the weighting scheme). All three specifications
show strong, highly significant positive spatial autocorrelation in the residuals,
motivating the spatial error model reported in Table S6.

\begin{table}[h]
\centering
\begin{tabular}{lcccc}
\toprule
Model & Moran's I & Expected I & z-score & p-value \\
\midrule
Baseline OLS             & 0.682 & $-$0.0001 & 72.93 & $<0.001$ \\
Interaction: Female       & 0.679 & $-$0.0001 & 72.60 & $<0.001$ \\
Interaction: Income       & 0.682 & $-$0.0001 & 72.91 & $<0.001$ \\
\bottomrule
\end{tabular}
\caption{Moran's I test for spatial autocorrelation in OLS residuals, using H3 hexagon contiguity weights.}
\label{si:tab:moran}
\end{table}

\end{appendices}
\end{document}